\documentclass[float,epsfig,usenatbib]{mn2e}

\usepackage{graphicx}
\usepackage{appendix}
\usepackage{psfrag}
\usepackage{amsmath,amssymb} 
\usepackage{threeparttable}
\usepackage{float}
\usepackage{subfigure}
\usepackage{afterpage}
\numberwithin{equation}{section} 

\title[ALFALFA H{\sc i} Data Stacking]{ ALFALFA H{\sc i} 
Data Stacking I. Does the Bulge Quench Ongoing Star Formation in
Early-Type Galaxies?} 
\author[S. Fabello et al.]
{Silvia Fabello$^{1}$\thanks{fabello@MPA-Garching.MPG.DE}, Barbara
  Catinella$^{1}$, Riccardo Giovanelli$^{2}$, Guinevere Kauffmann$^{1}$,
  \newauthor Martha P. Haynes$^{2}$,
  Timothy M. Heckman$^{3}$, David Schiminovich$^{4}$\\
$^{1}$Max-Planck Institut f\"{u}r Astrophysik, D-85741 Garching, Germany\\
$^{2}$Center for Radiophysics and Space Research, Cornell University, Ithaca, NY 14853, USA\\
$^{3}$Department of Physics and Astronomy, The Johns Hopkins
  University, Baltimore, MD 21218, USA\\
$^{4}$Department of Astronomy, Columbia University, New York, NY 10027, USA
}

\date{}

\begin{document}

\def\deg{$^{\circ} $}
\newcommand{\mone}{$^{-1}$}
\newcommand{\kms}{km~s$^{-1}$}
\newcommand{\kmsm}{km~s$^{-1}$~Mpc$^{-1}$}
\newcommand{\Ha}{$\rm H\alpha$}
\newcommand{\Hb}{$\rm H\beta$}
\newcommand{\hi}{{H{\sc i}}}
\newcommand{\hii}{{H{\sc ii}}}
\newcommand{\nii}{\ion{N}{2}}
\newcommand{\rband}{{\em r}-band}
\newcommand{\iband}{{\em I}-band}
\newcommand{\rd}{$r_{\rm d}$}
\newcommand{\whi}{$W_{50}$}
\newcommand{\x}{$\times$}
\newcommand{\about}{$\sim$}
\newcommand{\Msun}{M$_\odot$}
\newcommand{\Lsun}{L$_\odot$}
\newcommand{\Mhi}{M$_{\rm HI}$}
\newcommand{\Mst}{M$_\star$}
\newcommand{\must}{$\mu_\star$}
\newcommand{\col}{NUV$-r$}
\newcommand{\Rin}{$R_{50,r}$}
\newcommand{\Rout}{$R_{90,r}$}
\newcommand{\Rinz}{$R_{50,z}$}
\newcommand{\fgas}{$f_{\rm gas}$}
\newcommand{\sigc}{$\sigma_{1/8}$}
\newcommand{\cix}{$C$}
\newcommand{\Ra}{$\alpha$}
\newcommand{\dec}{$\delta$}

\maketitle

\label{firstpage}

\begin{abstract}\par
We have carried out an {\hi} stacking analysis of a volume-limited
sample of $\sim$ 5000 galaxies with imaging and spectroscopic data
from GALEX and the Sloan Digital Sky Survey, which lie within the current
footprint of the Arecibo Legacy Fast ALFA (ALFALFA) Survey.
Our galaxies are selected to have stellar masses greater than $10^{10} M_{\odot}$ and redshifts 
in the range $0.025<z<0.05$. We extract a sub-sample of 1833 ``early-type'' galaxies
with inclinations less than 70$^{\circ}$, with concentration indices $C>2.6$ and with light profiles
that are well fit by a De Vaucouleurs model. We then stack 
HI line spectra extracted from the ALFALFA data cubes at the 3-D positions of the  
galaxies from these two samples in bins of stellar mass, stellar mass surface density, 
central velocity dispersion, and NUV-$r$ colour.  We  use the stacked spectra 
to estimate the average {\hi} gas fractions {\Mhi}$/M_*$  of the galaxies in each bin.
Our main result is that the
{\hi} content of a galaxy is not influenced by its bulge. 
The average {\hi} gas fractions of galaxies in both our  samples                   
correlate most strongly  with  NUV-$r$ colour and with 
stellar surface density. The relation between
average {\hi} fraction and these two parameters is 
independent of concentration index $C$.  
We have tested whether the average {\hi} gas content
of bulge-dominated  galaxies on the red sequence, differs from that of
late-type galaxies on the red sequence.  We find no evidence that galaxies with
a significant bulge component are less efficient at turning their available gas reservoirs
into stars. This result is in contradiction with the  ``morphological quenching'' scenario
proposed by Martig et al. (2009).

\end{abstract}
 


\section{Introduction}

Galaxies have  long been  known to follow well-ordered  sequences in many  properties
\citep[see][for a review]{R&H}. In the simplest classification scheme,
galaxies divide into spirals and ellipticals. In more complex schemes, these classes are partitioned into additional
sub-classes. Irrespective of the exact classification scheme,  there are clear 
systematic trends in the bulge-to-disk ratios, surface brightnesses and in the  
concentration of light, all of which increase from spirals to ellipticals.  
Star formation rates and gas content decrease along this same sequence.

With the advent of large spectroscopic surveys of nearby galaxies, such as 
the Sloan Digital Sky Survey \citep[SDSS,][]{sdss}, the relationships between galaxy properties that
can be derived from the combination of optical imaging and spectroscopy, e.g. 
stellar mass, size, concentration index,  star formation rate, metallicity, dust content, have now
been systematized and  quantified in considerable detail   
\citep[e.g.,][]{kau03, Brich04, Trem04, bal04, bal06}. Our understanding of how the neutral gas content of galaxies relates to other galaxy  
properties lags far behind. 

The cold gas content of a galaxy  is known to vary strongly with  
colour and star formation rate.   
The  connection between gas content and galaxy  morphological type remains unclear. 
Whereas star-forming spiral galaxies almost always contain
{\hi} gas,  the {\hi} content of early-type galaxies is considerably more
difficult to predict.  Some Es and S0s have neutral atomic hydrogen gas content similar to those
of Sb-Sc type spirals, while others contain several orders of magnitude less {\hi} \citep{R&H}.  
It has been speculated that these variations may be an indication that the {\hi} gas in ellipticals
has an external origin \citep{Knapp85}, but direct proof of this conjecture is still lacking.        

Because the {\hi} is on average more difficult to detect in
  early-type galaxies, the samples discussed in the literature 
have generally been quite small. Some of the largest systematic studies of gas in early-type
galaxies were carried out in the 1980's. \citet{Knapp85}
analyzed a sample of 152 nearby elliptical  galaxies, of which 23 were
detected in the {\hi}
line. These authors studied the distribution of the quantity 
 {\Mhi}/{L$_B$} for these systems. In contrast to spiral galaxies,
 where the distribution N({\Mhi}/{L$_B$}) has a well-defined mean value and a small dispersion, {\Mhi}/{L$_B$} 
spans a wide range in ellipticals.  \citet{Wardle86} extended this work 
to S0 galaxies and found that the {\hi} detection rate was twice as high
compared to  ellipticals.
Some years later,  \citet{Breg92} carried out a study of the interstellar components
of 467 early-type galaxies in the Revised Shapley Ames Catalogue and again reported a trend of
increasing neutral gas content from E to Sa.
These authors  suggested that the cold gas in early-types  is associated
mainly with  disks and not with the bulge components of these galaxies.
However, more recent studies that have mapped {\hi} in nearby early-type galaxies \citep{mor06}  
have  concluded that the {\hi}
can be organized in a variety of different configurations, e.g. 
in regular disks, in clouds, in rings,  or even in tidal tail-like structures. 

One major problem that has plagued our understanding of cold gas in early-type galaxies
is that the available {\hi} data have been  inhomogeneous.   
Large area, blind {\hi} surveys such as the Arecibo Legacy Fast ALFA
Survey \citep[ALFALFA;][]{alfalfa} offer uniform coverage over large regions of the sky and allow
one to construct complete, unbiased samples of {\hi}-selected galaxies. However, these
surveys are shallow, and do not in general detect gas-poor early-type
galaxies.
This limitation has been pointed out in recent papers by   \citet{diS08} and \citet{Grossi09}, which used
ALFALFA data to study an unbiased sample of early-type galaxies in the
Virgo cluster  region. They were able to compare the {\hi} content of early-type galaxies drawn from field and
group environments, but their average detection rates were much smaller than earlier
studies based on incomplete and inhomogeneous data. 

The Hubble morphological classification scheme is based on the optical
appearance of a galaxy which in turn depends both on structural
properties such as its bulge-to-disk ratio, and on  star formation  rate.
If we wish to understand the  physical processes that regulate the neutral  gas content 
of galaxies, it is preferable to analyze  the effects of  star formation and  galaxy structure
{\em separately}.
Recently, \citet{Helm07} studied 30 E and S0  galaxies with
signs of recent or ongoing star formation  
and concluded that such systems are more gas rich than E/S0 galaxies with old stellar populations.
The availability of 
sizes, surface brightnesses and  parameters measuring the concentration of the light for samples 
of millions of galaxies made possible due to recent advances in
large scale CCD surveys and automatic image processing techniques, enable a new approach
to understanding the interplay between stars and gas in early-type systems.

In this paper, we make use of stacking techniques to analyze whether the average {\hi} gas fraction
of a galaxy is affected by the presence of a significant bulge
component.  
Stacking has now become a common tool to constrain the statistical properties of 
a population of objects that lack individual detections in a survey; 
by co-adding the signal from many objects with known sky positions and redshifts,  
the background noise can be decreased and one can recover the average {\hi} flux of the ensemble.
Stacking techniques have been applied to a wide variety of different astrophysical data. Examples include 
studies of faint radio AGN \citep{Hodge09}, studies of star
formation in high redshift Lyman Break Galaxies \citep{Car08},
studies of the intracluster light using stacked  optical images \citep[e.g.,][]{Zib05}, 
and the soft X-ray properties of high redshift quasars
\citep{Shen06}. 
Stacking has been applied to {\hi} data as well, with the purpose
of studying the {\hi} properties of gas at redshifts that are
currently not well probed by existing radio telescopes \citep[individual
detections of {\hi} emission reach redshifts z$\sim$0.25 and require
extremely long integrations,][]{Cat08}. \citet{Cheng01} stacked non-detections in
different regions of a z=0.06 cluster to investigate environmental
effects. More recently, {\hi} stacking was used by \citet{ver07} to
probe the Butcher-Oelmer effect at z$\sim$0.2, and by \citet{Lah07,
  Lah09} to attempt to constrain the {\hi} content of star-forming
galaxies at z$\sim$0.24 and galaxies around a cluster at z$\sim$0.37.

Here, we  use ALFALFA survey data to constrain the average {\hi} gas fractions  of an
unbiased sample of massive early-type galaxies.  We study how the {\hi} content depends on
parameters such as  stellar mass, stellar mass surface
density, concentration index, central velocity dispersion and UV/optical colour. 
The paper is structured as follows. In section \ref{sample} we
describe the samples considered in this paper. The stacking analysis 
is described in section \ref{stacktool}. First we study 
{\hi} gas fraction scaling relations for a complete sample of galaxies
and then we compare
our  results with previous work (section
\ref{stackGass}). The analysis of the {\hi} properties of early-type galaxies  
is presented in section \ref{etg}. 
All  distance-dependent quantities in this work are computed
assuming $\Omega_m\,=\,$0.3, $\Lambda\,=\,$0.7 and $H_0\,=\,$70 km$\,\,$s$^{-1}\,$Mpc$^{-1}$.

\section{sample selection}\label{sample}
\begin {figure*}
\begin {tabular}{c}
\includegraphics[width=14.cm]{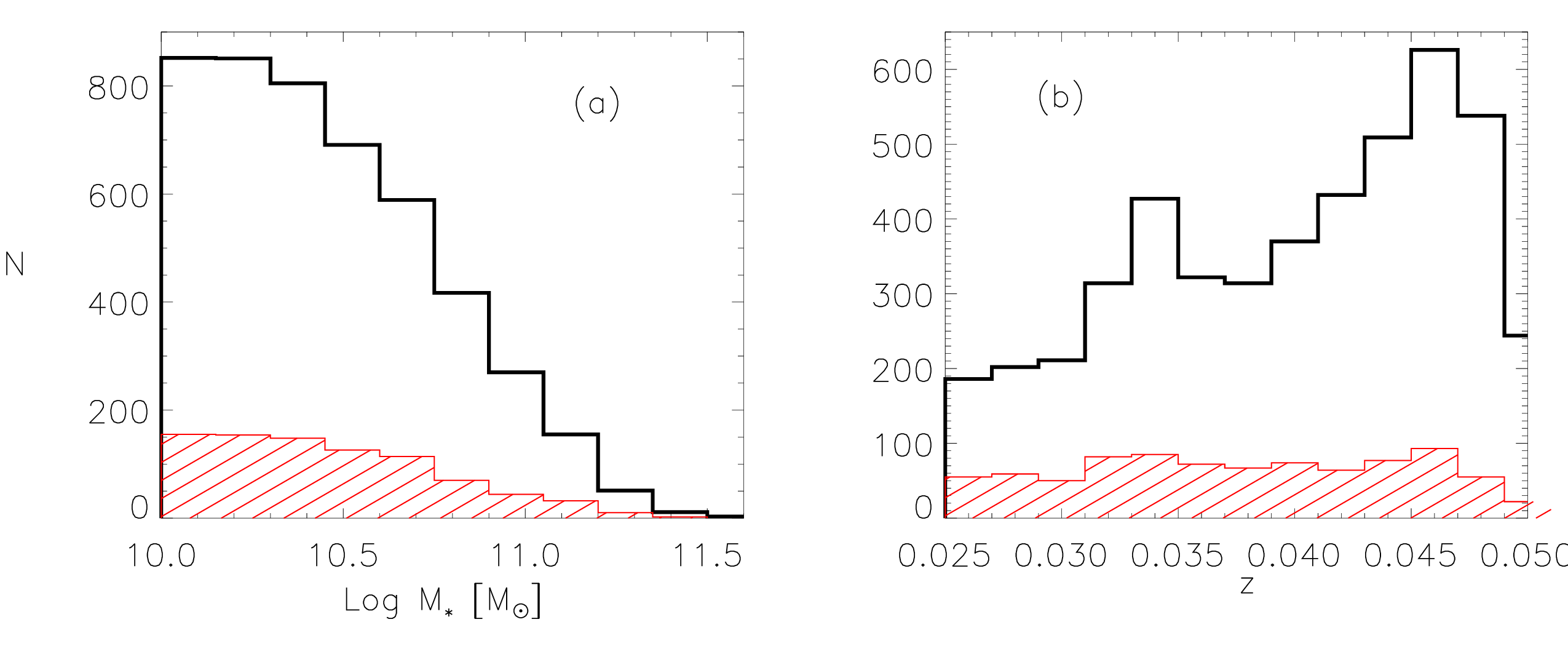}\\
\includegraphics[width=14cm]{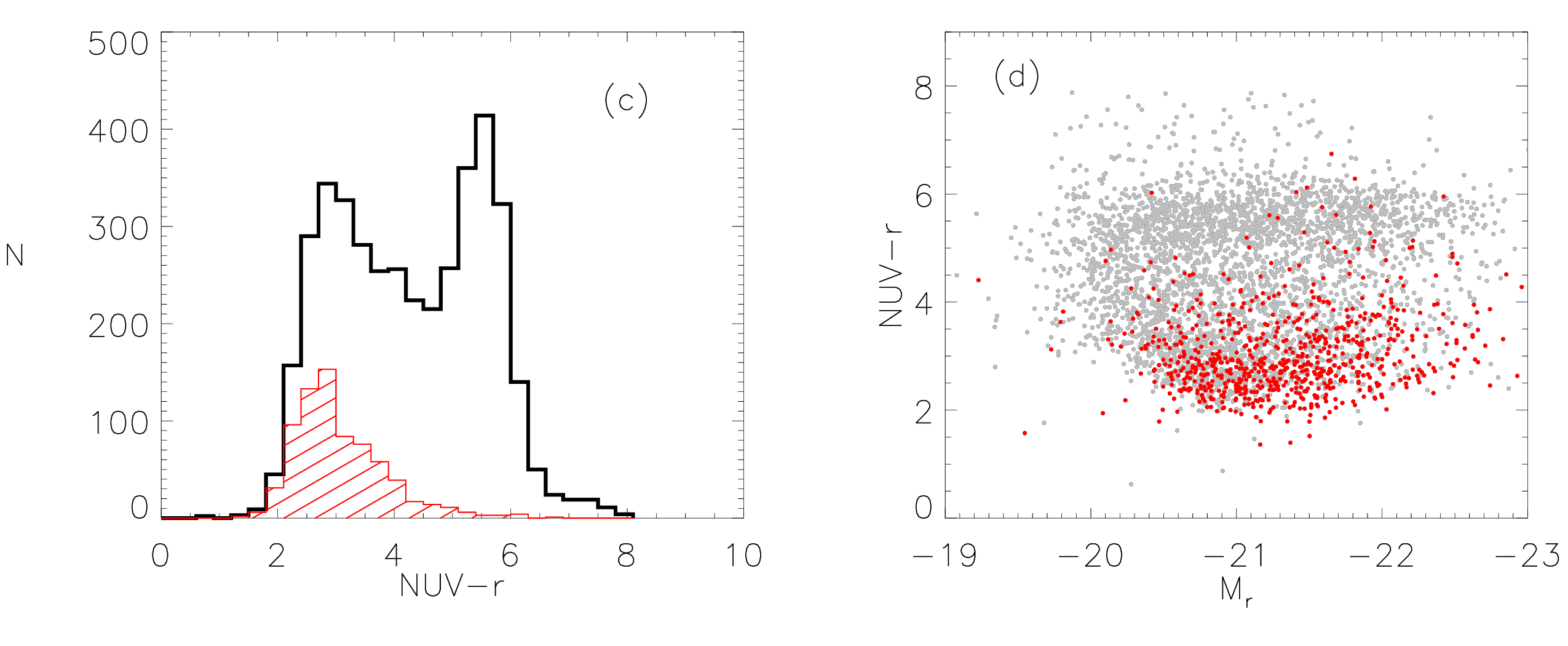}
\end {tabular}\caption{ Distributions of (a)
stellar mass, (b) redshift and (c) NUV-$r$ colour (corrected for Galactic
 extinction only) for galaxies in \textit{sample A}. The black solid histograms
represent the whole \textit{sample A}, while the red, dashed histograms show the distributions
for the sub-sample of galaxies with ALFALFA detections. Panel (d) shows the colour-magnitude
diagram of \textit{sample A} galaxies (gray dots) and the sub-sample with
ALFALFA detections (red dots).}    
\label{fig01}\end{figure*} 
Measured redshifts are essential if we wish to recover an accurate estimate
of the mean {\hi} content of a population of galaxies using stacking techniques.  
Our sample is drawn from the ``parent sample'' of the GALEX Arecibo SDSS Survey \citep[][hereafter GASS-1]{gass01}, which is a 
sample of 12006  galaxies with stellar masses greater than  $10^{10} M_{\odot}$
and redshifts in the range $0.025<z<0.05$ selected from the
SDSS main spectroscopic sample. The parent sample galaxies are
located in the intersection of the footprints of the Data Release 6
of the SDSS \citep[DR6;][]{DR6}, the projected GALEX Medium Imaging
Survey \citep[MIS;][]{galex} and ALFALFA.                                   
The average uncertainty in the SDSS spectroscopic redshifts
is extremely small (0.0002).
As  discussed in GASS-1 (see section 3 for further details), the stellar masses are derived from SDSS
photometry following \citet{salim07} and have  typical errors
smaller than 30\%. The choice of redshift range is determined  by sensitivity limits in the {\hi}
observations and by the need to avoid redshift ranges where radio frequency interference (RFI)
is a problem.

The acquisition of ALFALFA data is on-going. Part of the data have
already been catalogued and are available to the public \citep[for the
ALFALFA sky covered by SDSS:][]{aadr1,aadr3,aadr5}. 
The ALFALFA 40\% dataset to be released in late 2010 (Martin {\it et al.} in preparation;
Giovanelli {\it et al.}, in preparation)
includes the following SDSS sky regions: $7.5\,
\mathrm{h}<\alpha_{2000}<16.5\,
\mathrm{h},~+4^{\circ}<\delta_{2000}<+16^{\circ}$ and
$+24^{\circ}<\delta_{2000}<+28^{\circ}$, and $22\,
\mathrm{h}<\alpha_{2000}<3\,
\mathrm{h},~+14^{\circ}<\delta_{2000}<+16^{\circ}$.\par
Within the same sky region, the
GASS sample contains 5350 galaxies; these constitute the parent
sample for our study.
We discard some objects because they have poor quality
{\hi} data (see section
\ref{stacktool_a}), so that the final GASS-SDSS-ALFALFA
sample (which we call \textit{sample A}) is composed of 4726 objects. 
Of these, 23\% are galaxies with reliable ALFALFA detections (i.e. objects corresponding to ALFALFA 
detection codes 1 or 2\footnote{As discussed in \citet[]{alfalfa}, ALFALFA
HI line detections are coded into two categories: 
  Code 1 detections have a peak signal-to-noise ratio greater than 6.5 and are reliable
  at greater than 95\% confidence; Code 2 detections, referred to as ``priors'', have a lower 
  signal-to-noise ratio between 4.5 and 6.5 but an optical
  counterpart at the same known redshift. Their reliability
  is estimated to be greater than 85\%.}).
Panels (a), (b) and (c) of Figure \ref{fig01} show the stellar mass, redshift and NUV-$r$ colour distributions
of the galaxies in \textit{sample A}. 
The NUV-r colours have been  corrected for foreground Galactic extinction, but not for  internal extinction.
The ALFALFA detection rate is close to 23\% for each stellar mass
  bin in Figure \ref{fig01}, but is clearly biased to  blue-sequence
  objects (with {\col}$\,\lesssim\,$3.5) as shown in panel (c).  
Finally, in panel (d) we plot NUV-$r$ colour versus absolute $r$-band magnitude $M_r$ for the sample,
with black dots representing the full sample and red points the galaxies detected by ALFALFA. 
Once again we see that the galaxies detected by ALFALFA are almost exclusively found on the blue sequence.

\subsection {Galaxy Parameters}
\begin {figure*}
\begin {tabular}{c}
\includegraphics[width=14cm]{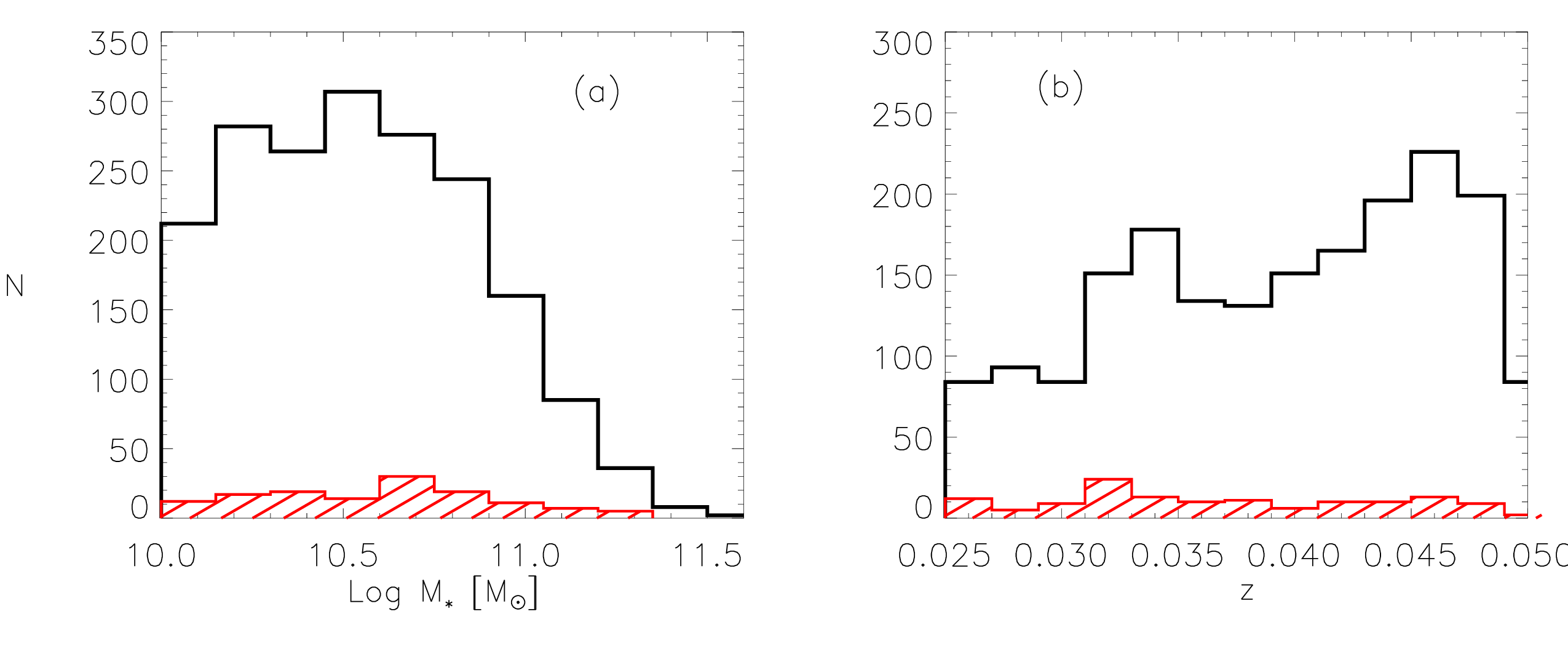}\\
\includegraphics[width=14cm]{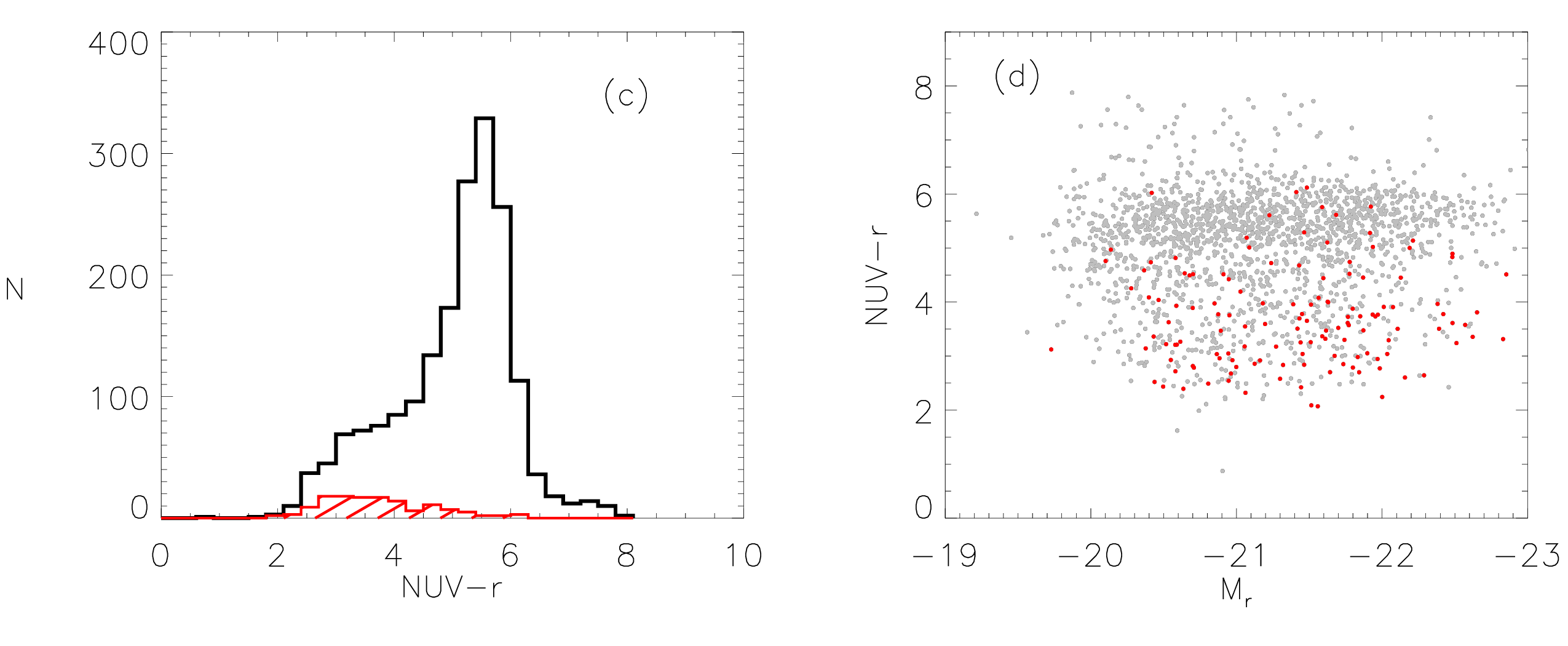}
\end {tabular}\caption{ As in Figure 1, except for the \textit{ETG sample}.}
\label{fig02}\end{figure*} 
The optical parameters we use are derived from the MPA-JHU SDSS DR7
release of spectrum measurements or from Structured Query Language
(SQL) queries to the SDSS DR7 database
server\footnote{See http://www.mpa-garching.mpg.de/SDSS/DR7/ and http://cas.sdss.org./dr7/en/tools/search/sql.asp}.
The UV parameters are extracted from the GALEX UV photometry by \citet{Jing09}.  
The reader is referred to section 5 of GASS-1 for more
detailed descriptions.

The parameters  used in this paper are the following:
1) stellar mass {\Mst}, 2) stellar mass surface density {\must}, defined as
{\must}={\Mst}/(2$\pi R^2_{50,z}$), where $R_{50,z}$ is the Petrosian
radius containing 50\% of the flux in $z$-band in units of kpc,
3) the concentration
index {\cix}=$R_{90}/R_{50}$, where $R_{90}$
and $R_{50}$ are the radii enclosing 90\% and 50\% of the $r$-band
Petrosian flux, 4) and NUV-$r$ colour. As already explained, the colour is corrected for Galactic
extinction only. Corrections for internal dust-attenuation are
discussed and applied in \citet{gass2}, where they study the star formation
properties of the GASS sample. \\

In addition to these quantities, we extract the following photometric parameters:
 \begin{enumerate}
\item The inclination $i$ to the
line-of-sight is evaluated according to: $\cos{i}=b/a$, where $a$
and $b$ are the semi-major and semi-minor axes from the  $r$-band
 exponential fit, respectively ($b/a$ for the exponential fit is tabulated as \textit{expAB\_R}).  
\item The likelihood parameters (\textit{lnldev\_r} and
   \textit{lnlexp\_r} from SDSS), which indicate how well the de
   Vaucouleurs and the exponential models fit the one-dimensional  $r$-band
   light profile of the galaxy
\item  The central velocity
 dispersion $\sigma_{1/8}$ is derived from the SDSS  parameter
\textit{vdisp}. These velocities are evaluated with the ``direct
fitting'' method\footnote{See  
http://www.sdss.org/DR6/algorithms/veldisp.html for more discussion} using
spectra measured within the 3$''$-diameter fiber aperture. 
Only values between 70 and 420 km$\,$s$^{-1}$
are reliable. 
We then correct \textit{vdisp} for aperture effects. 
Following \citet{Graves09a} and references therein, we scale the
fiber velocity dispersion to that at 1/8 of
the effective radius: $\sigma_{1/8}\,=\,vdisp\cdot \left(r_{fib}/\frac 1
8 r_{\circ}\right)^{0.04}$, where $r_{fib}=1''.5$ and $r_{\circ}$
is the circular galaxy radius, defined as $r_{\circ} = R_e\sqrt{b/a}$.
$R_e$ is the r-band de Vaucouleurs radius (tabulated as \textit{devR\_r}), and $a$,
$b$ are now the major and minor axis from the de Vaucouleurs fit
($b/a$ is tabulated as \textit{devab\_r}). In general, this correction 
is small, $\sim 5$\%.
(Note that for galaxies included in  \textit{sample A}, $R_e$ has a mean value of 8.2$''$.)  
\end{enumerate}

\subsection{ETG sample selection}\label{etg_sample}
In this paper, we have chosen to define ``early-type'' galaxies purely in terms
of their structural properties, without regard to their stellar populations or
star formation rates. We note that our definition is in contrast to some definitions
of ``early-types''  in the literature, which have excluded galaxies with emission lines 
\citep[e.g.,][]{Bernardi03, Graves09a}. Our goal will be to explore the extent to 
which the presence or absence of a significant bulge component influences the {\hi}
content of a galaxy, so we do not wish to bias our conclusions by selecting against
bulge-dominated galaxies with emission lines.

Starting from \textit{sample A}, we extracted a subset of
early-type objects (\textit{ETG sample}) with the following
properties: 
\begin{itemize}
 \item concentration index {\cix}=$R_{90}/R_{50}\,\ge
   2.6$;
 \item the likelihood that the light profile is fitted by a de
   Vaucouleurs model is greater than it is by an exponential one;
 \item inclination less  than 70$^{\circ}$ (this cut rejects the most inclined systems).  
 \end{itemize}

We note that the {\cix} parameter has
been shown to be an excellent indicator of the bulge-to-total
  ratio (B/T) derived from full 2-dimension bulge/disk decomposition
analysis \citep{Gad09}. 
A cut at  {\cix}$\,\ge\,$2.6 restricts the sample to galaxies with
$B/T \ge\, 0.4$ \citep[see Fig.1 of][]{Wein08}. 
For the present work we choose our default cut at {\cix}$\,\ge\,$2.6, but we
also experiment with cuts at larger values of {\cix}.  

\begin {figure*}
\includegraphics[width=16cm]{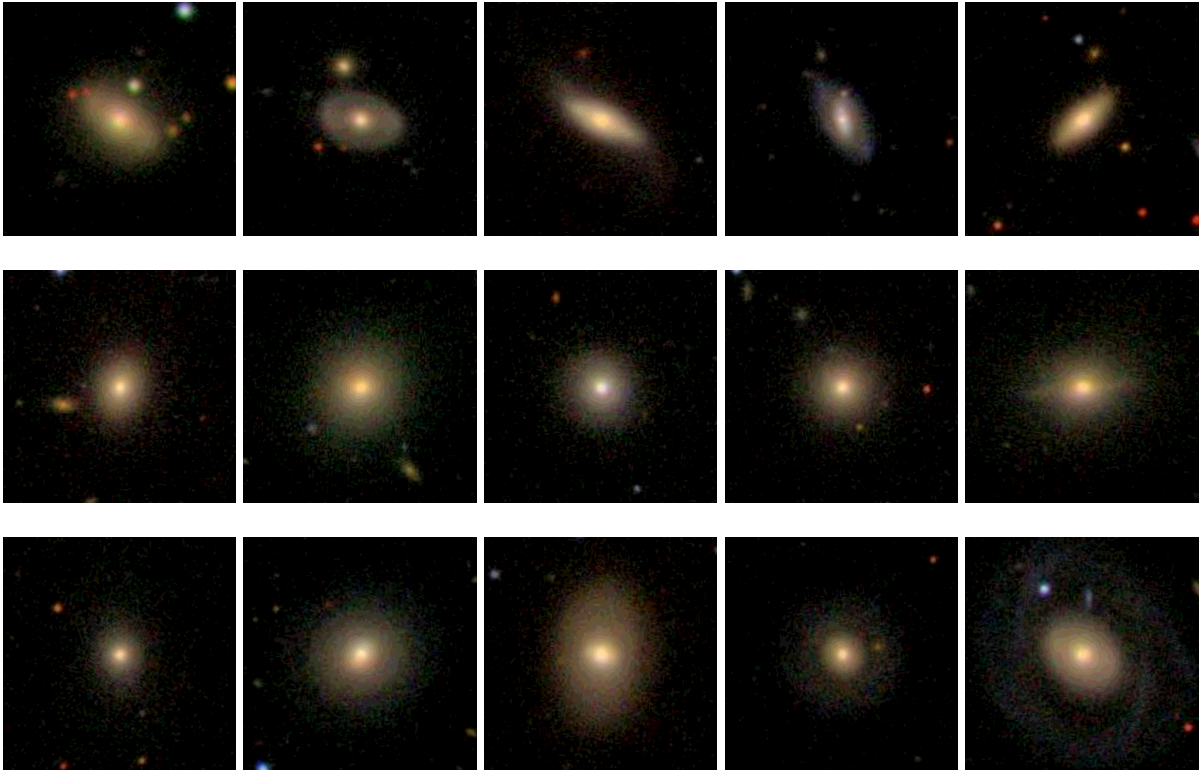}
\caption{SDSS postage images of galaxies randomly selected from
    the \textit{ETG sample} and ordered by increasing {\cix}. From left to
    right, top to bottom {\cix} increases from 2.6 till 3.8. The images are 1 arcminute square in size. }\label{fig03}
\end{figure*}

Application of our default cut leads to a final \textit{ETG sample} consisting of 1833 objects.
The properties of this sample are shown in Figure
\ref{fig02}. The solid histograms represent the
entire sample; the red dashed histograms the
sub-sample detected by ALFALFA. The stellar
mass, redshift and the NUV-$r$ colour 
distributions are shown in panels
(a), (b) and (c), respectively. The average ALFALFA detection rate  for early-type objects
is smaller ($\sim$ 9\%) than for \textit{sample A}.
The colour-magnitude diagram is shown in panel (d). Most of the
early-type targets lie on  a well-defined red sequence. 
Some objects do scatter bluewards of the red sequence. 
Such objects may be star-forming,
transitional or Seyfert galaxies  \citep{schaw07}. By selecting targets
based on concentration index and inclination, we may also include
 objects with some disk component. In Figure
 \ref{fig03} we presents 1$\times$1 arcminute SDSS postage stamps of a
 randomly selected set of galaxies from our \textit{ETG sample}, that
 have been ordered by increasing {\cix}.
\begin {figure*}
\begin {tabular}{cc}
\includegraphics[width=8cm]{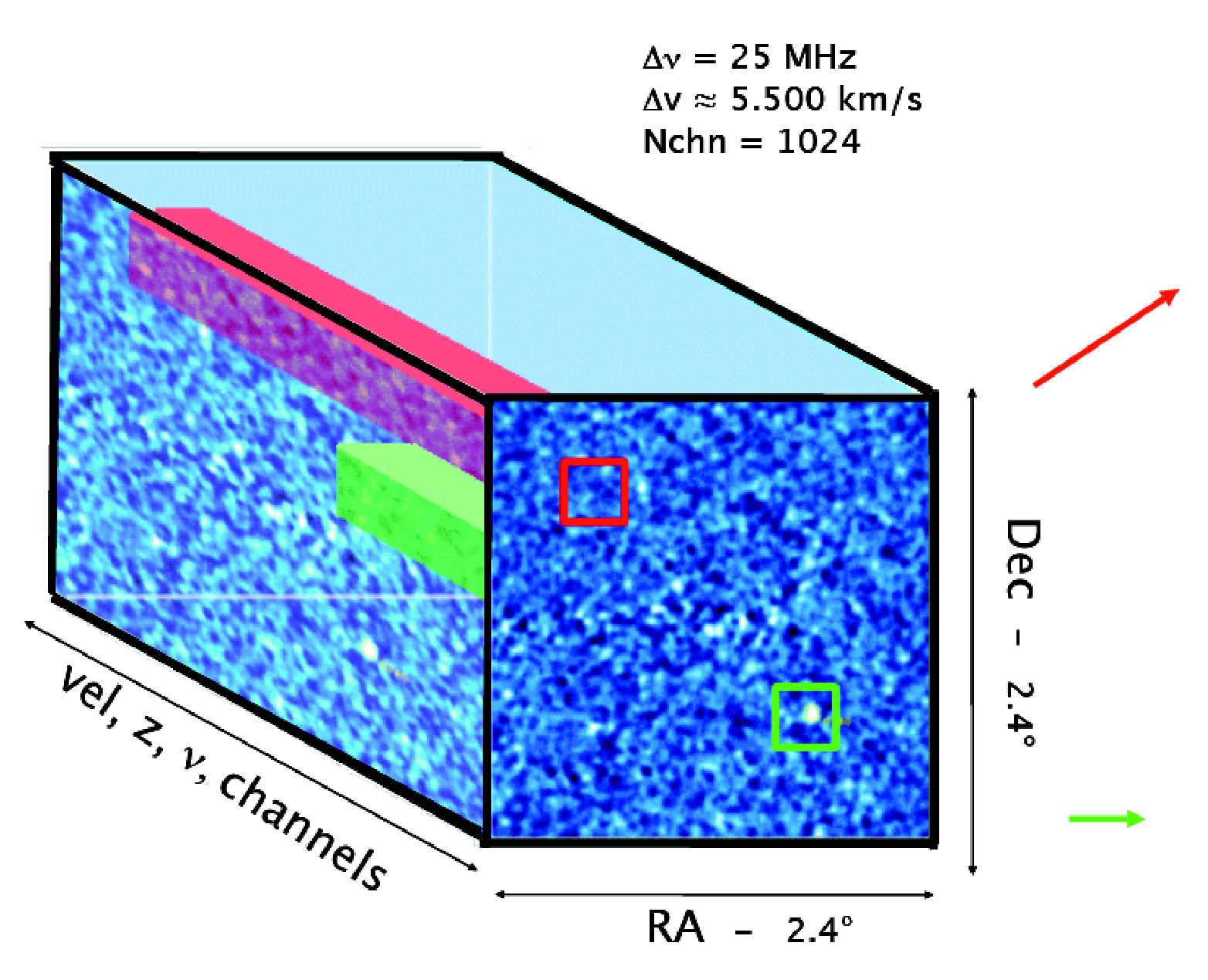} & 
\includegraphics[width=6cm]{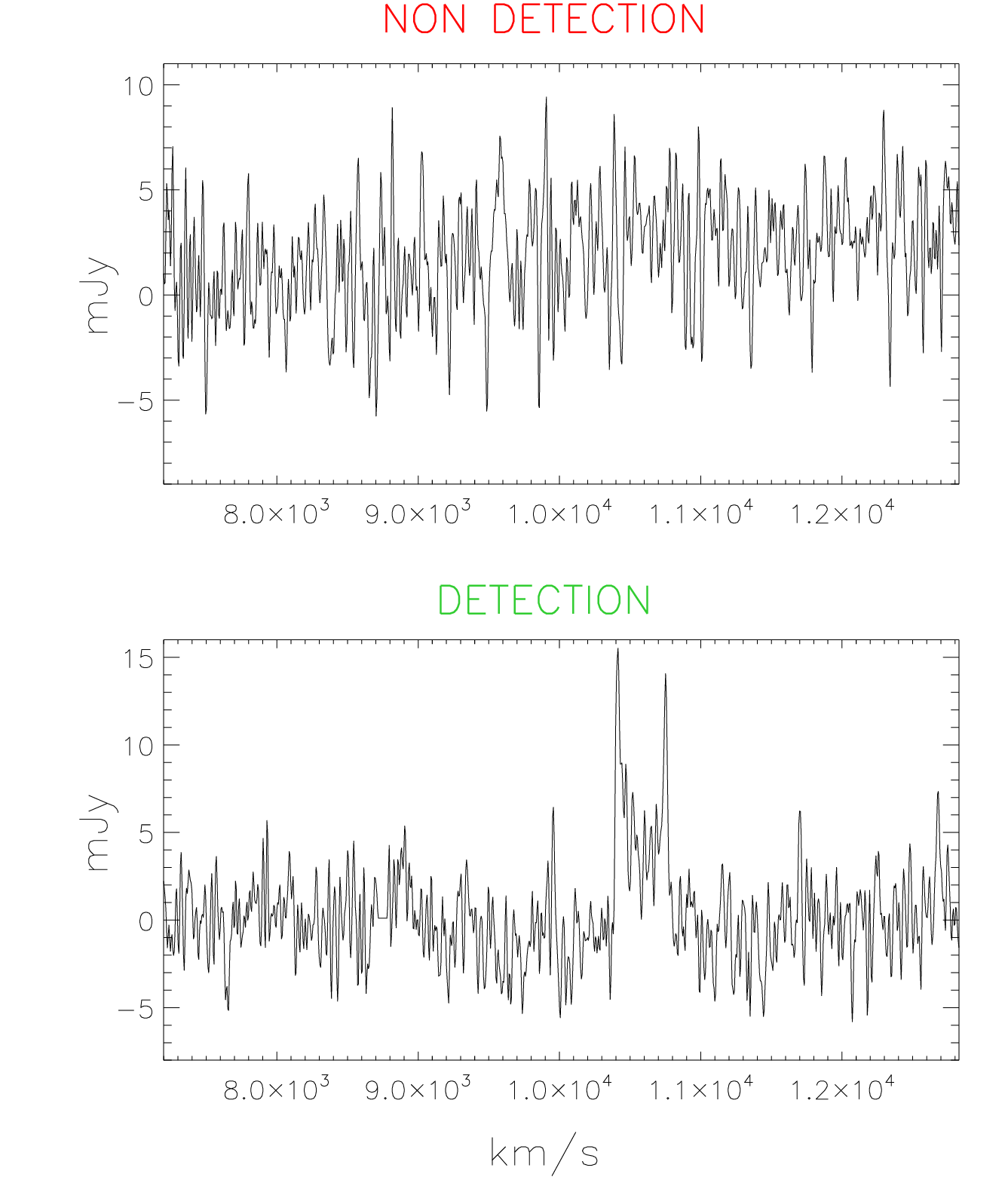}
\end {tabular}\caption{A schematic representation of the
fully processed ALFALFA 3-D data-cube. The data cubes are 2.4$^{\circ}\times$2.4$^{\circ}$ in size
and about 5500 $\mathrm{km}\,\mathrm{s}^{-1}$ in velocity range
(25 MHz in frequency). The raw spectral resolution is $\sim$5.5
$\mathrm{km}\,\mathrm{s}^{-1}$; the angular resolution is $\sim$4$'$. 
For each pixel, which is a point in RA, Dec and velocity, a value of flux density is recorded. 
For each target in \textit{sample A}, we extract a spectrum at a given position of the sky, over the 
velocity range of the data-cube which contains the source. Two
examples of extracted spectra are shown on the right,  illustrating an {\hi}
detection (green, bottom) and an {\hi} non-detection (red, top).}\label{fig04}\end{figure*} 

\section{ALFALFA data stacking}\label{stacktool}

The on-going ALFALFA survey is scanning 7000 deg$^2$ of the high galactic latitude
sky over the velocity interval
$v\mathrm{[km\,s^{-1}]}\,\simeq\,$[-2500;18000] (i.e. out to  z$\sim$0.06)
using a simple ``minimum-intrusion'' drift scanning technique\citep[]{alfalfa}
that exploits the seven-horn Arecibo L-band feed array (ALFA). For each of the
seven beams, spectra are recorded separately for the two orthogonal polarizations,
providing two independent samples.
The 2-D drift scan data from both polarizations and all beams 
covering a given portion of the sky are combined
to form 3-D cubes of dimension  
2.4$^{\circ}\times$2.4$^{\circ}$ on the sky and 
5500 $\mathrm{km}\,\mathrm{s}^{-1}$  in velocity ``depth". 
We refer the reader to Figure
\ref{fig04}, where a schematic representation of one
such data-cube is shown. We will refer to this depiction throughout the paper.

The raw spectral resolution of the ALFALFA data, before smoothing, is $\sim$5.5
$\mathrm{km}\,\mathrm{s}^{-1}$ and the angular resolution is $\sim$3.3$'
\times$3.8$'$ (corresponding to the FWHM of each ALFA beam). 
The data-cubes or ``grids'' are constructed from the drift scan data so that
each spatial pixel is 1$'$ on a side. The 2-D drift scans are flux calibrated
using the real-time noise diode injection scheme; the position and flux scales
of the final grids are updated using fits to the many radio continuum
sources they contain \citep[]{aadr3}. The ALFALFA processing scheme \citep[]{alfalfa}
retains all individually recorded spectra for each ALFA beam and each of its
polarizations separately; no filtering is performed to discard bad channels. 
In order to deal with poor quality data, the 2-D data for each beam and polarization
are visually inspected to flag bad records and frequency channels. 
The 3-D grid construction then proceeds
with knowledge of the flagged 2-D pixels. Data quality in the final grid then
may be limited by the absence of data in channels flagged as contaminated by RFI, 
weighted as poor by occasional instrumental problems or missing entirely because
of incomplete sky coverage. In order to account for these effects, each 3-D pixel of the  data-cube is
assigned a quality weight $w$ during the data reduction process, which is a number ranging from 
0 (unusable data) to 20 (good data), computed according
to the availability and quality of data contributing to each pixel. The 
retention of the full dataset and the construction of the accompany ``weights map'' allows us to 
judge whether or not adequate data exist for each target so that it can be meaningfully
included in the stacking process.

\subsection{Creating a catalogue of {\hi} spectra}\label{stacktool_a}
We extract a spectrum for each galaxy in the
sample. As discussed, all our targets are selected from the SDSS
spectroscopic survey, so we know both their
positions on the sky  and their redshifts. We first  select the ALFALFA
data-cube which contains the target and then follow a procedure that includes the following steps:
$a)$ spectrum extraction; $b)$ rms evaluation, $c)$
final quality check. \\

{\bf a. Spectrum extraction}\\ 
The signal from each  target is integrated over a region of the data cube
centered on its 3-D position. Because noise increases with the square root of 
the integration area, integrating over too large a region lowers 
the quality of the spectrum without increasing the signal. Our     
GASS targets are always smaller then the ALFA beam (the mean $R_{90}$
for \textit{sample A} is 10$''$), so we simply integrate over a sky
region  of 4$'\times$4$'$. In Figure
\ref{fig04}, we illustrate how we extract spectra at two different
positions in the sky inside the same data-cube. The coloured regions indicate  where the
spectra would be evaluated.  

The {\hi} spectrum is a histogram of flux
density $S$ as a function of velocity. For each velocity channel $v$, the corresponding flux density $S_v$ is obtained by
integrating the signal $s_v(x,y)$ over the spatial pixels centered at the target galaxy
position, as observed by a radio telescope of beam response pattern B:
\begin{eqnarray*}
 S_{v}\mathrm{[mJy]}=\frac{\Sigma_{x}\Sigma_{y}
  s_{v}(x,y)}{\Sigma_{x}\Sigma_{y}
   B(x,y)},\end{eqnarray*}
where $x$, $y$ are the sky coordinates (the two polarizations are kept
separated). The expression above means that the spatially integrated
profile is obtained by summing the signals over all the spatial pixels
of interest and dividing by the sum of the normalized beam $B(x,y)$ over the same
pixels \citep[for a detailed discussion see][]{sho79}. The ALFALFA
beam pattern can be approximated by: 
\begin{eqnarray*}
B(x,y)=\exp\left(-\frac 1 2 \left(\frac {x}
  {\sigma_x}\right)^2-\frac 1 2 \left(\frac {y}
  {\sigma_y}\right)^2\right),\end{eqnarray*}
with  $\sigma_x=(2\sqrt{2\ln 2})^{-1}\times 3.3'$, and $\sigma_y=(2\sqrt{2\ln 2})^{-1}\times
3.8'$ \citep[]{alfalfa}.   

We note that we discard any spectrum if more than  40\% of the pixels have a
a quality weight $w$ less than 10.
We also keep track of the three strongest continuum sources in an area covering 
40$'\times$40$'$ around each source; strong continuum sources  can affect our spectra by
creating standing waves\footnote {Standing waves are periodic fluctuations in the
background which occur when radiation from a strong continuum source
is multiply reflected and scattered by the telescope structure before reaching the receiver.}.\\

{\bf b. Rms evaluation}\\
For each spectrum we need to measure the root mean square (\textit{rms})
noise, which will later be used
as a weighting factor when we stack spectra. The \textit{rms} has to be evaluated
in regions of the spectrum where there is no emission from the target galaxy, and we also have to avoid
spectral regions where there are any spurious signals (e.g. residual RFI that we failed to flag, or 
{\hi} emission from companion galaxies). In order to define the spectral region that might contain galaxy
emission, we estimate its expected {\hi} width as follows. \\
The expected  width of the {\hi} spectrum will depend on the  rotational
velocity of the galaxy  as observed along the line-of-sight. We estimate the expected
velocity  $w_{TF;o}$ for each target, using the the Tully-Fisher
relation. Following  \citet{TF97b}, we use the SDSS $i$-band magnitude
($k$-corrected and corrected for Galactic and internal extinction as
in equations 11 and 12 in \citet{TF97a}) to estimate $w_{TF}$, and the measured inclination of the galaxy 
to derive  
 $w_{TF;o}$. We are aware that the Tully-Fisher relation does not hold for all morphological
types and environments. We do not think this is a major issue, because these velocities are only used to 
estimate  the region of the spectrum that should contain significant signal from the galaxy.  

We then fit a first order polynomial to the
baseline after  excluding the  region of the spectrum containing signal from the galaxy.
This step allows us to eliminate possible gradients
in the background. (The top right panel in Figure  \ref{fig04} shows an example of a spectrum    
where the baseline is tilted). We
perform a robust polynomial fit over the  regions of the spectrum  with high
values of the quality factor $w$ and then evaluate the    
\textit{rms} about the fit over the same region of the spectrum.

The average \textit{rms} for the whole sample (for each polarization) is
3.6$\,\pm\,$0.5 mJy. After averaging the two polarizations, an \textit{rms} of
$\sim$2.5 mJy is obtained; this is comparable to the average
\textit{rms} of 2.2 mJy evaluated for published, reduced ALFALFA
spectra.\\

 {\bf c. Final quality check}\\
After we have extracted the spectra, we visually inspect each of
them. We check the extraction process, and we discard spectra            
with bad baselines caused by continuum sources and those with
possible spurious signals close to the galaxy (e.g., if one polarization
has a significantly stronger signal than the other, or if there is a strong
signal close to the object, which may arise from a companion galaxy). 
These cuts eliminate 624 objects in  \textit{sample A}  
(11\% of the initial sample). 

\subsection{The Stacking Method}\label{stacktool_b}
We want to co-add the signals from $N$ different sources located at different
redshifts. First we shift each spectrum to the galaxy rest
frequency, so each spectrum is centered at zero velocity. 
We stack together the spectra $S_i$ (i=1,..N) 
using their \textit{rms} as a weight, a
standard approach in stacking analysis. In doing this, the final spectrum
$S_{stack}$ would be:
\begin{eqnarray}\label{eq:stack}
S_{stack}&=&\frac{\Sigma_{i=0}^{N}\, 
   S_{i}\cdot w_{i}}{ \Sigma_{i=0}^{N}\,w_{i}}\\
w_{i}&=&\frac{1}{\textit{rms}_{i}^2}\:\:.\end{eqnarray}
The stacking of the spectra is done separately for each polarization.
Note that ALFALFA is a blind survey which scans the
sky uniformly, so the  \textit{rms} for most spectra are similar.
If the noise of the input targets is purely Gaussian, the \textit{rms} of the
stacked spectrum is expected to decrease as 1/$\sqrt{N}$, where $N$
is the number of objects co-added. In reality, in addition to the Gaussian
noise there are likely to be systematic components,
for example standing waves (we
discarded spectra dominated by standing waves, but a weak residual signal could
still remain). Because of these additional noise sources, we expect
the  \textit{rms} to approach a lower 
limit as the number of co-added spectra becomes 
very large.  
 \begin {figure}
\includegraphics[width=8.5cm]{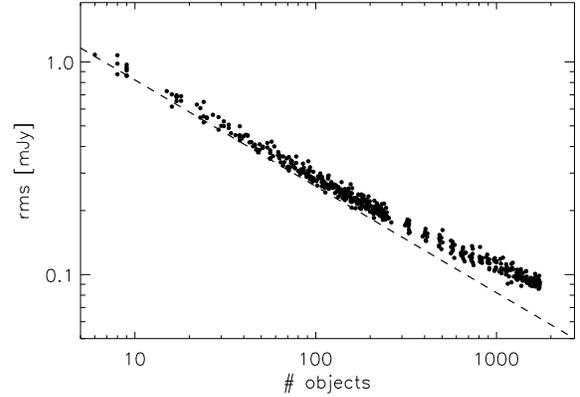}
\caption{ Dependence of the \textit{rms} of the stacked spectra as a function of
the number of objects co-added. The dashed line is the
expected 1/$\sqrt{N}$ dependence. The flattening of the relation around $N\,\sim \,$300
arises, as expected, where the non-Gaussian noise becomes dominant.}
\label{fig05}\end{figure} 
 
In Figure \ref{fig05} we show how the measured \textit{rms} of the stacked
spectra decreases as a function of the number of co-added objects. 
We have stacked increasing numbers of randomly selected
galaxies from  \textit{sample A},
and for each stack we  evaluated an \textit{rms} as described
above. These measurements are shown as dots, 
while the dashed line shows the expected average \textit{rms} value assuming Gaussian
noise, i.e. 
$2.5/\sqrt{N}$ mJy. As expected, we see the
trend flatten as N approaches values of around 300, where
the non-Gaussian noise becomes dominant. The \textit{rms} continues
to decline as N increases, but at a slower rate.  

After stacking, we manually process each ensemble spectrum. The reduction process
includes the following steps: \textit{i)} we average the two polarizations. 
\textit{ii)}
A default Hanning smoothing is performed on the spectrum.
Depending on the  signal-to-noise,  we may apply a boxcar smoothing to
decrease the noise if the signal is marginal. Note that
we  never average over more than nine
channels (corresponding to $\Delta v \sim
50\,$km$\,$s$^{-1}$). \textit{iii)} Finally, the baseline is subtracted.
We note that the  baseline of a stacked spectrum is already almost flat, because
the different noise features tend to cancel
out when averaging many spectra.

\begin {figure*}
\begin {tabular}{cc}
\includegraphics[width=7.4cm]{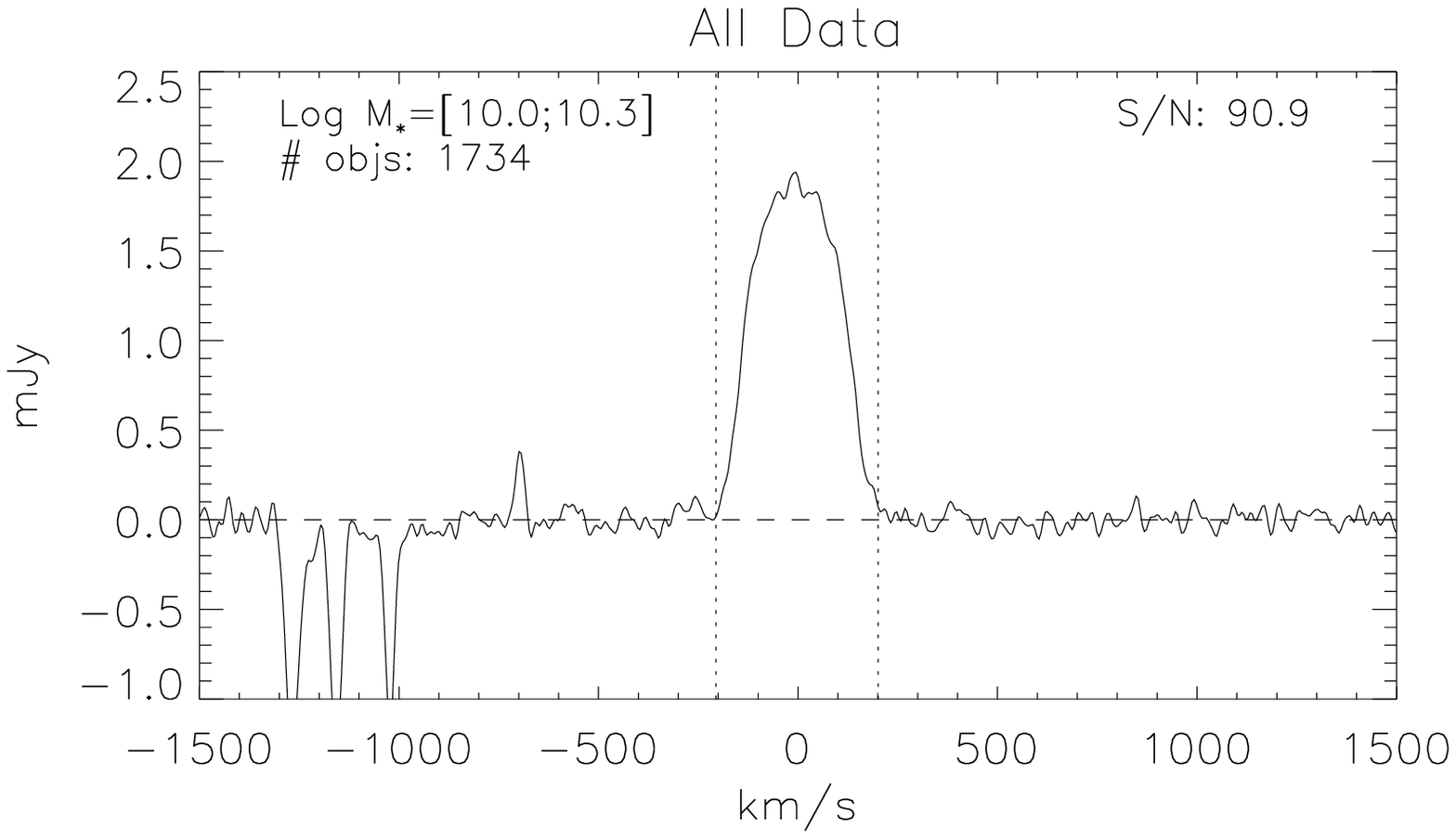}
&\includegraphics[width=7.4cm]{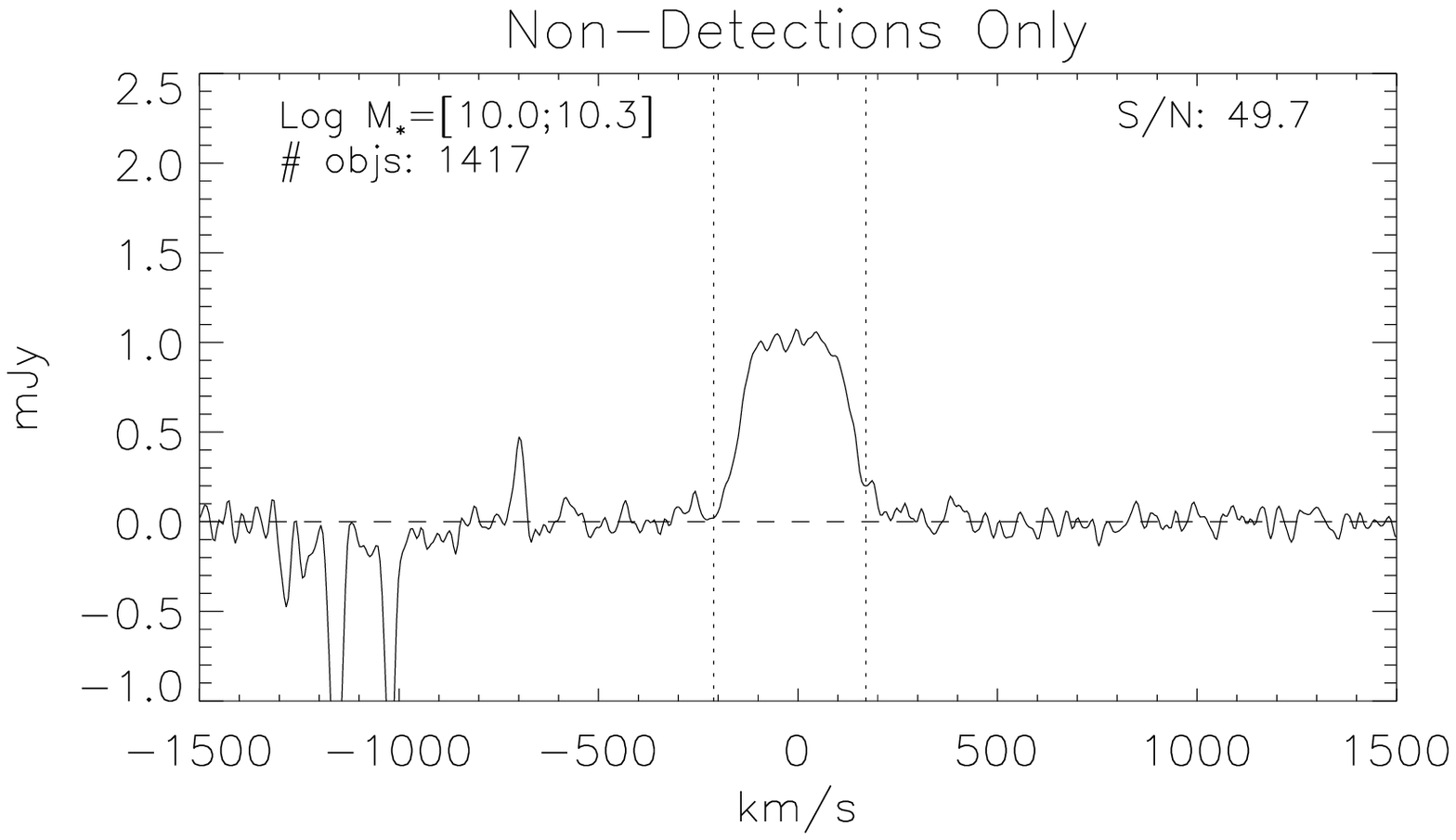} \\
\includegraphics[width=7.4cm]{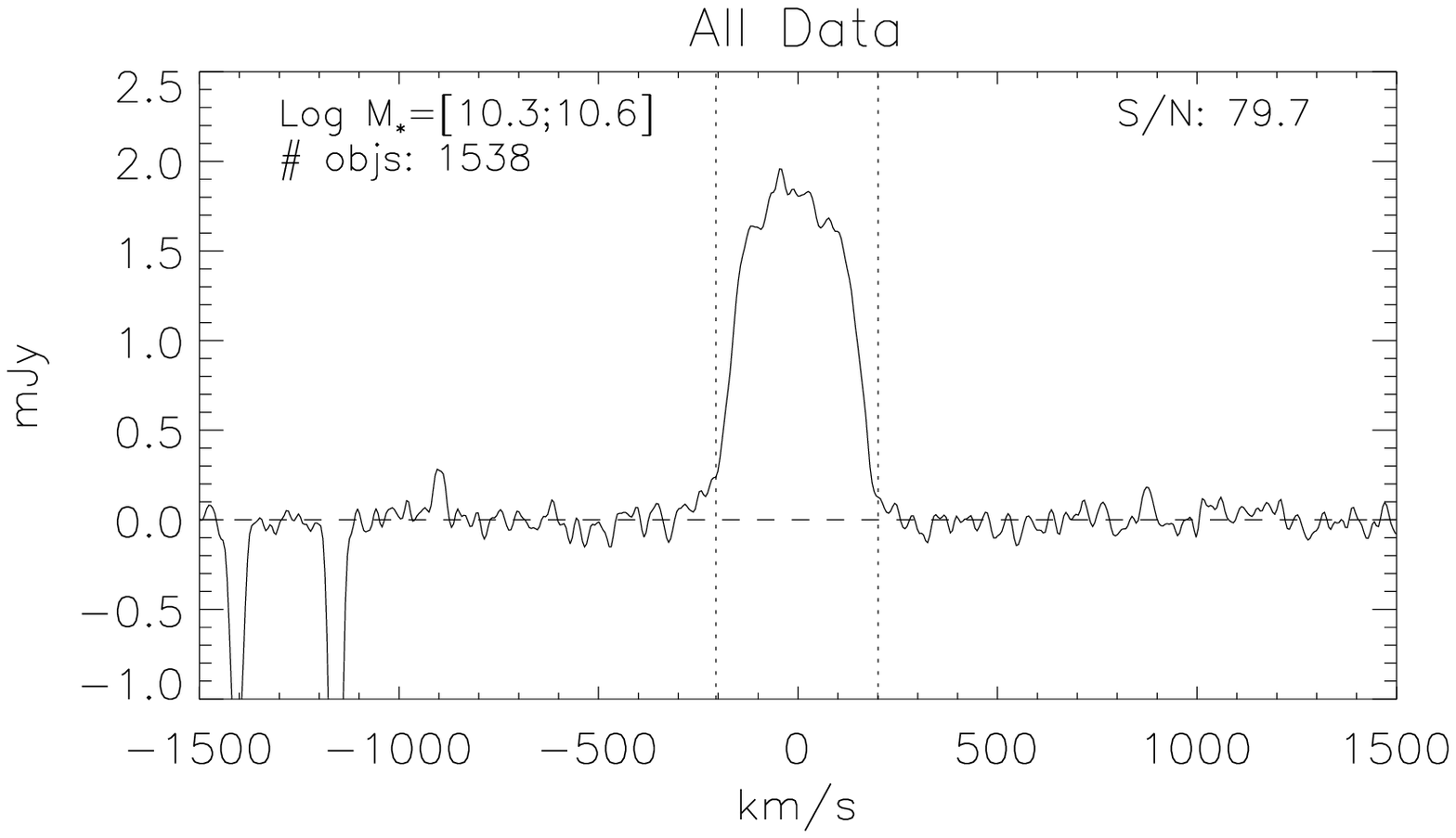} 
&\includegraphics[width=7.4cm]{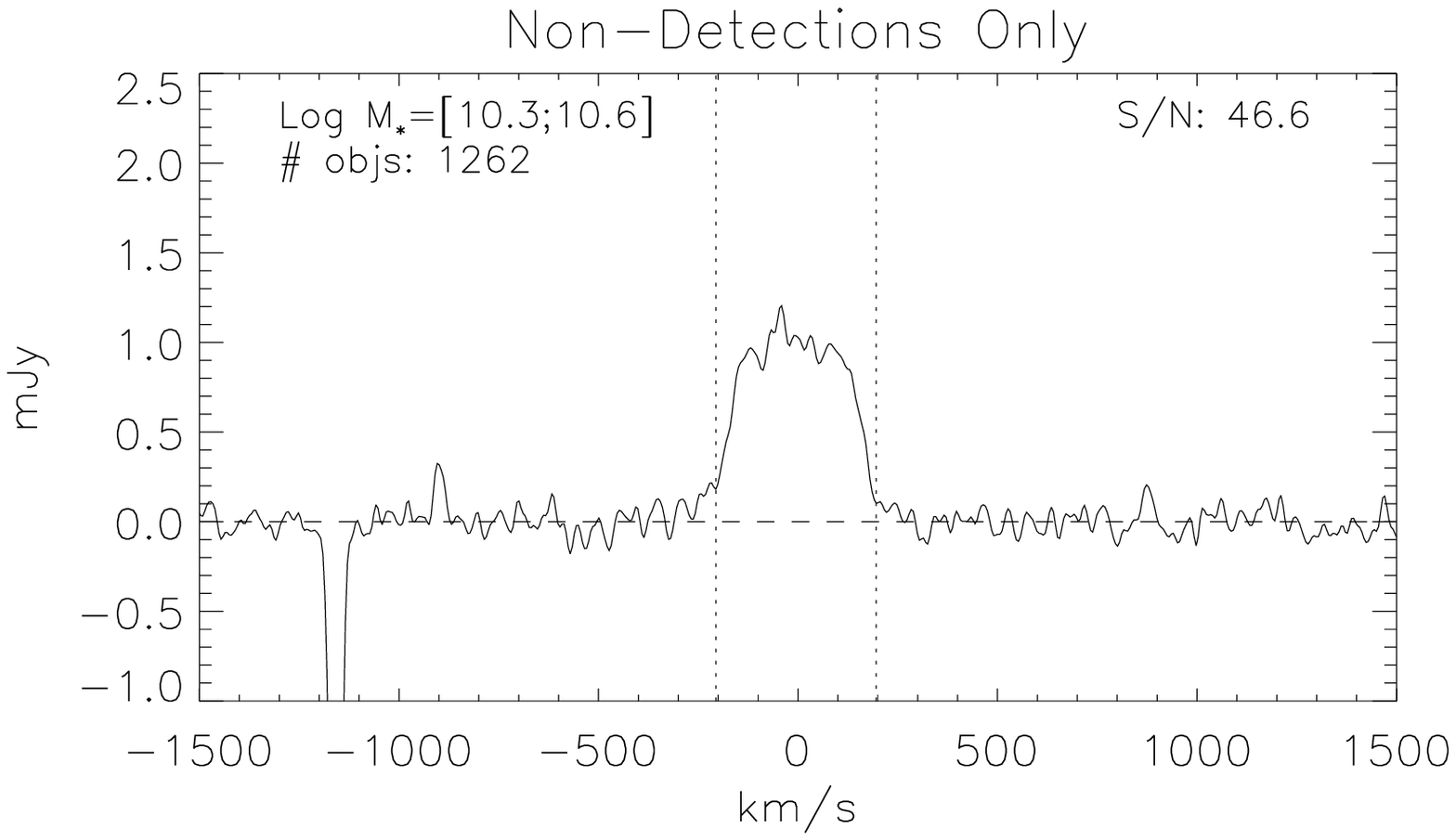} \\
\includegraphics[width=7.4cm]{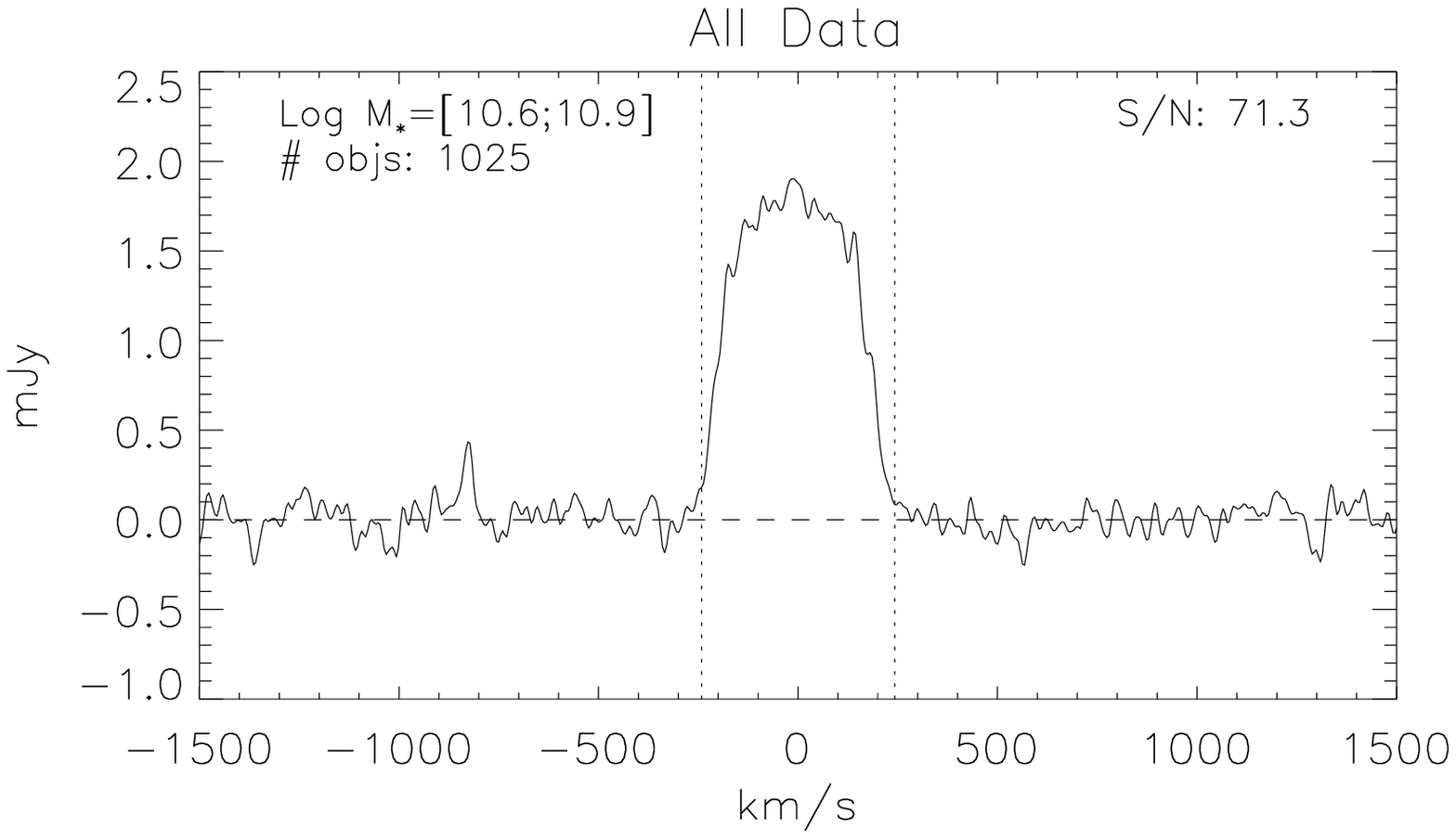}
&\includegraphics[width=7.4cm]{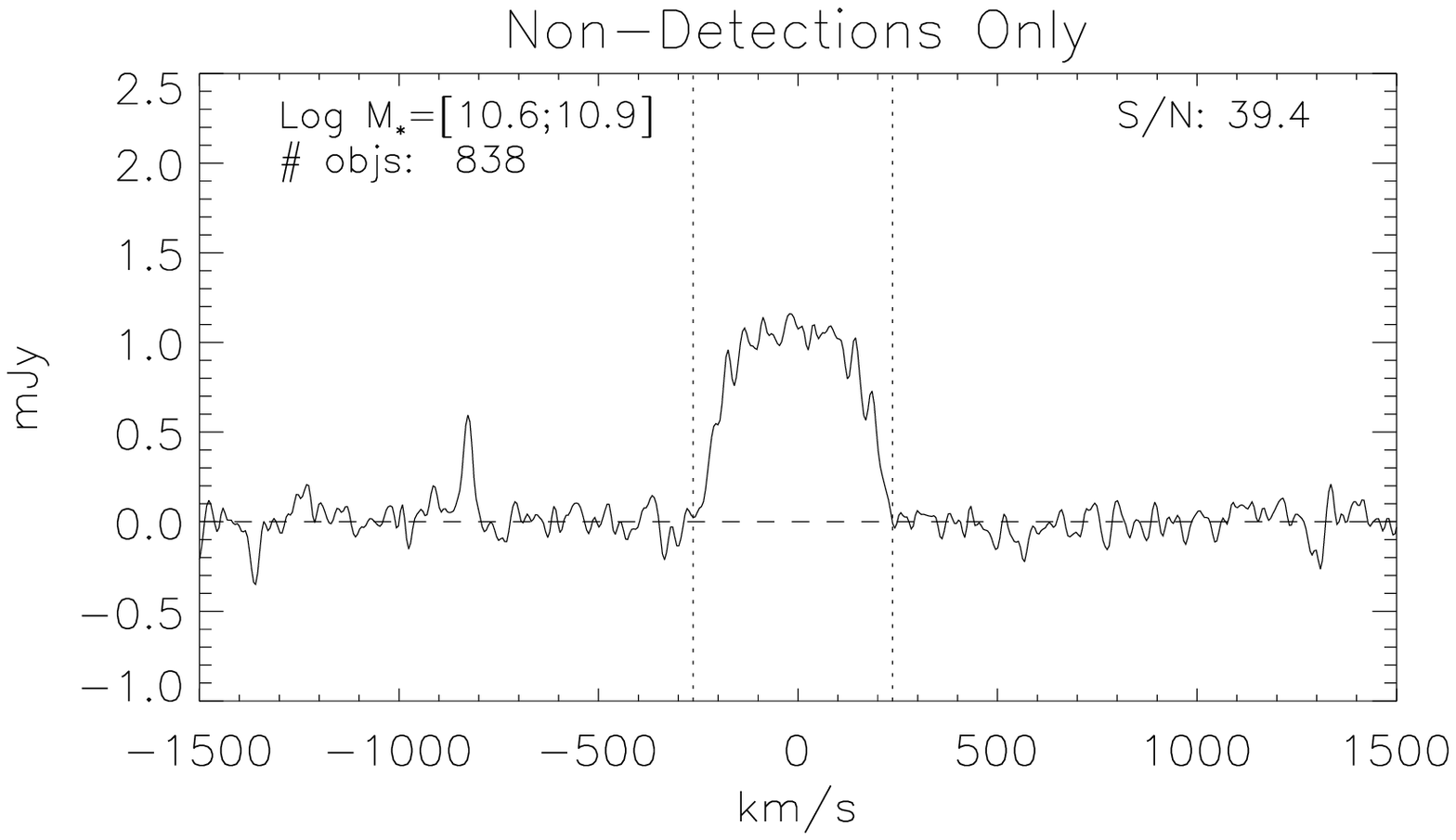}\\
\includegraphics[width=7.4cm]{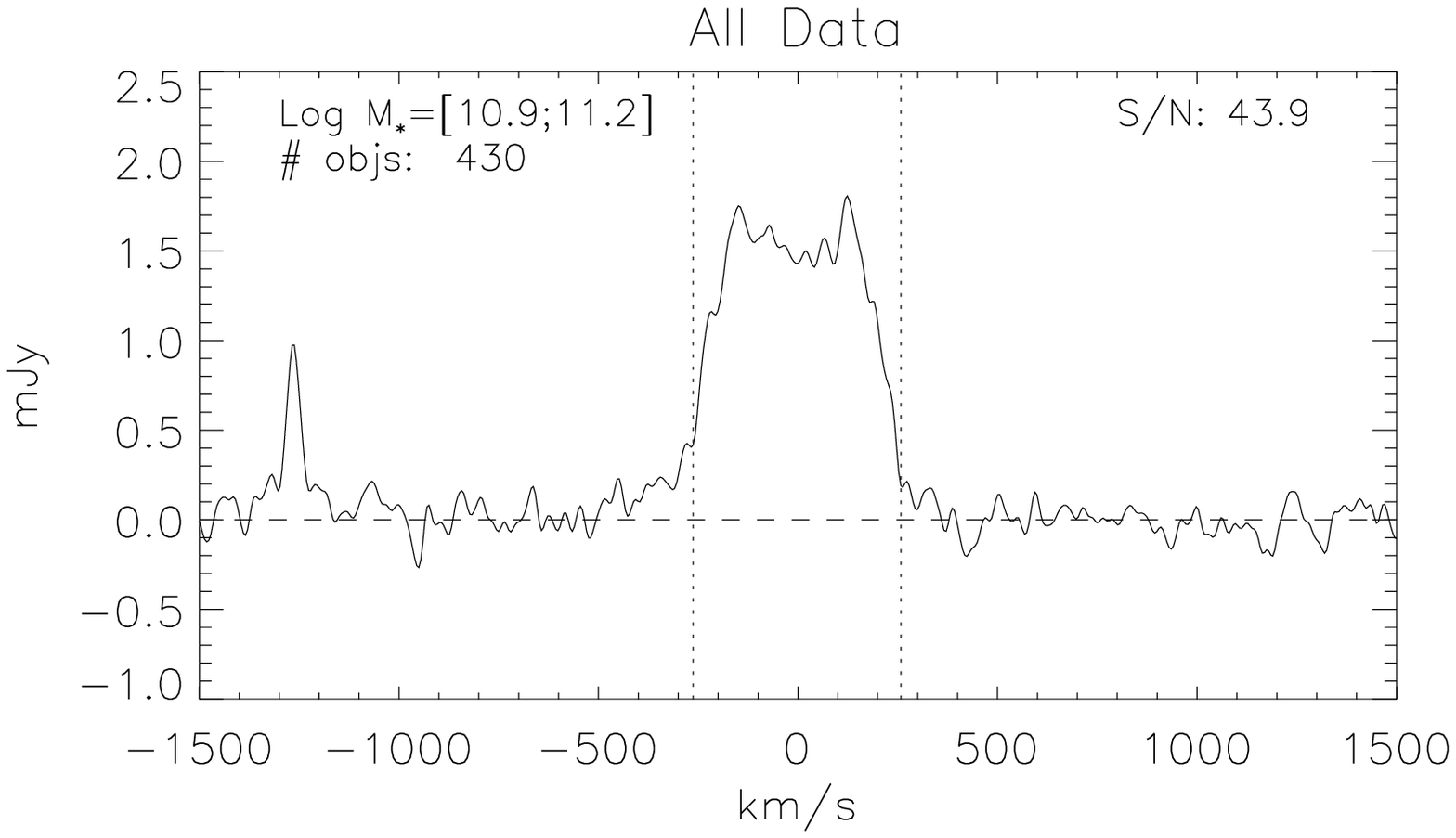} 
&\includegraphics[width=7.4cm]{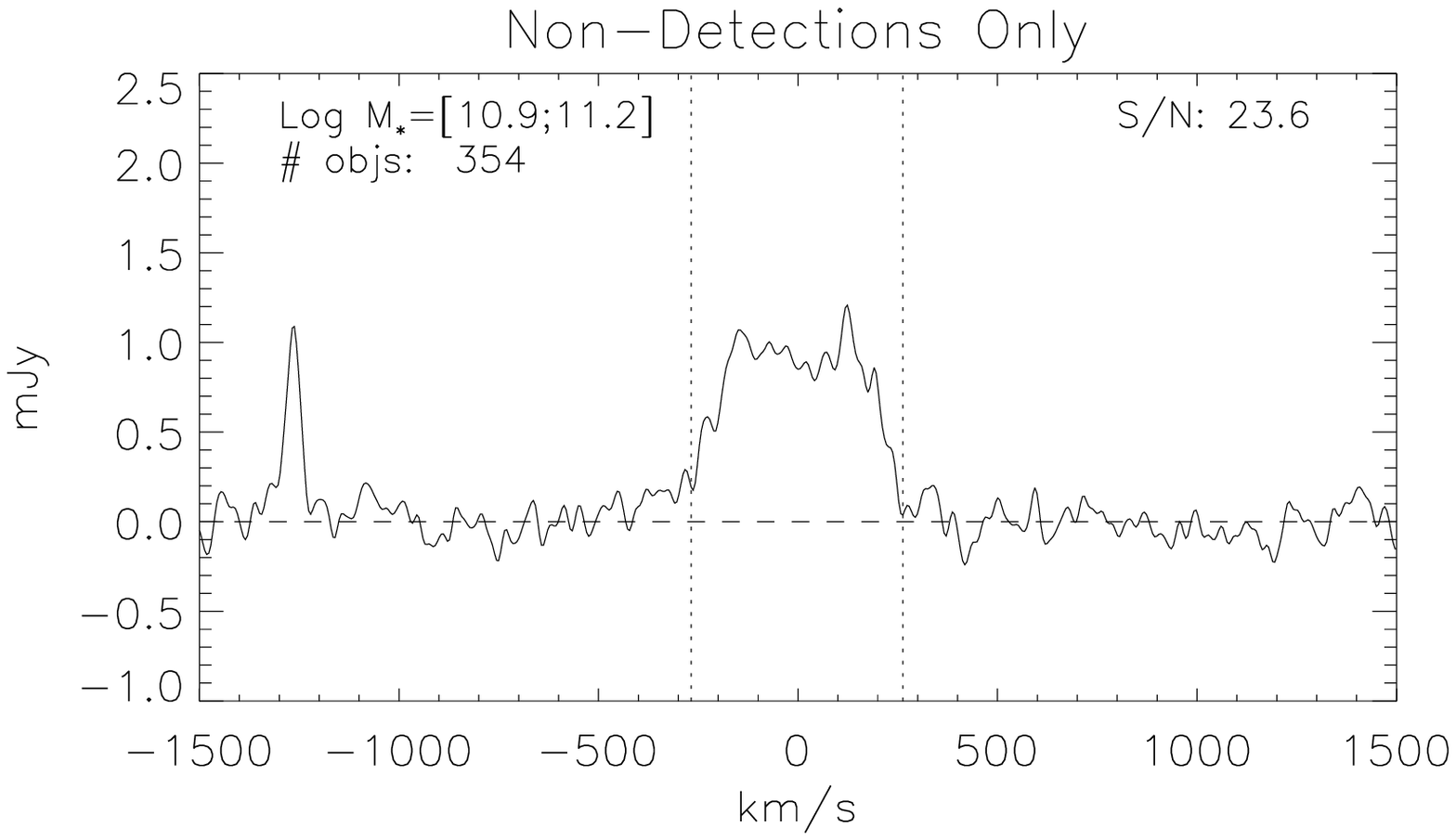} \\
\includegraphics[width=7.4cm]{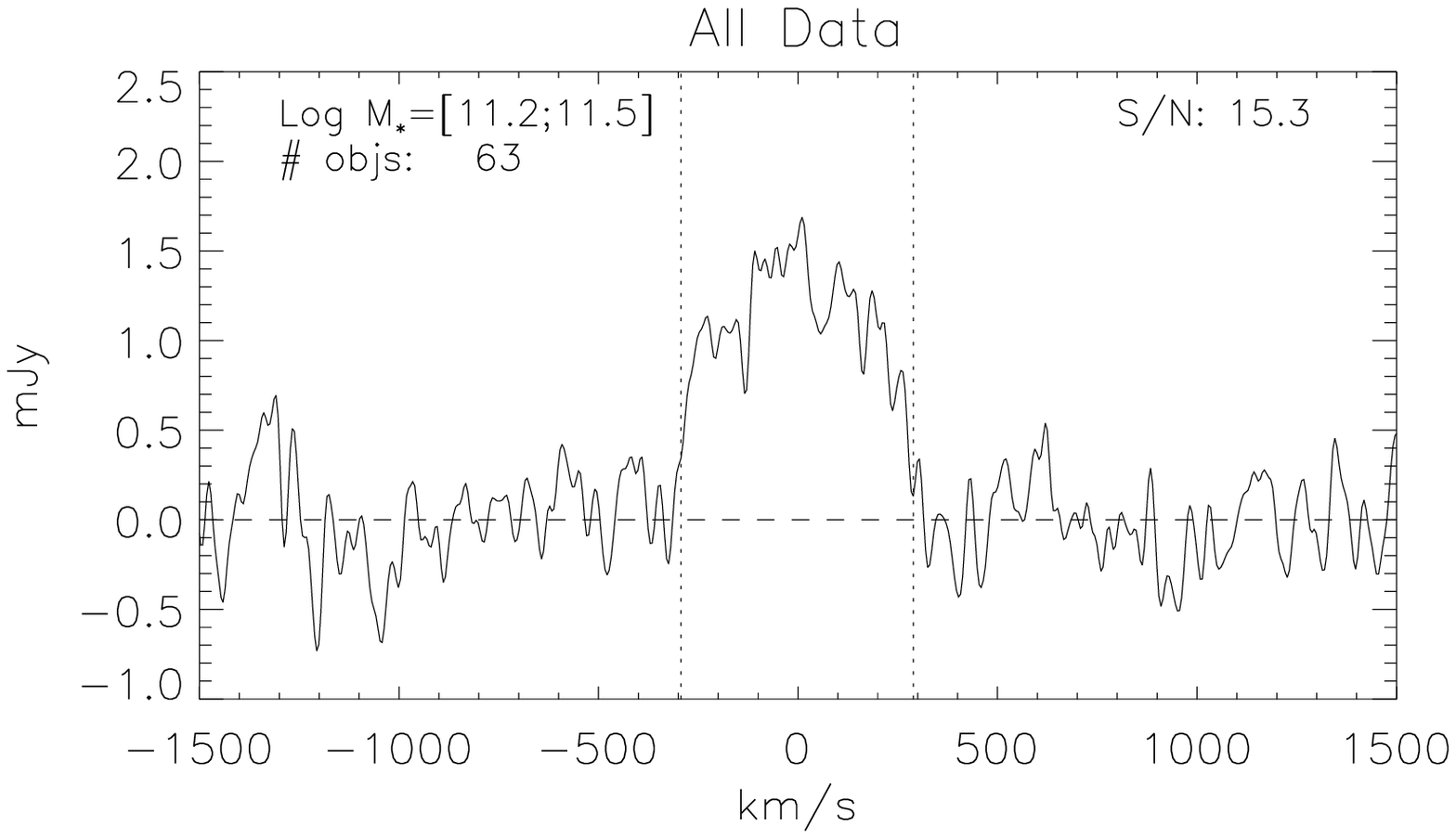} 
&\includegraphics[width=7.4cm]{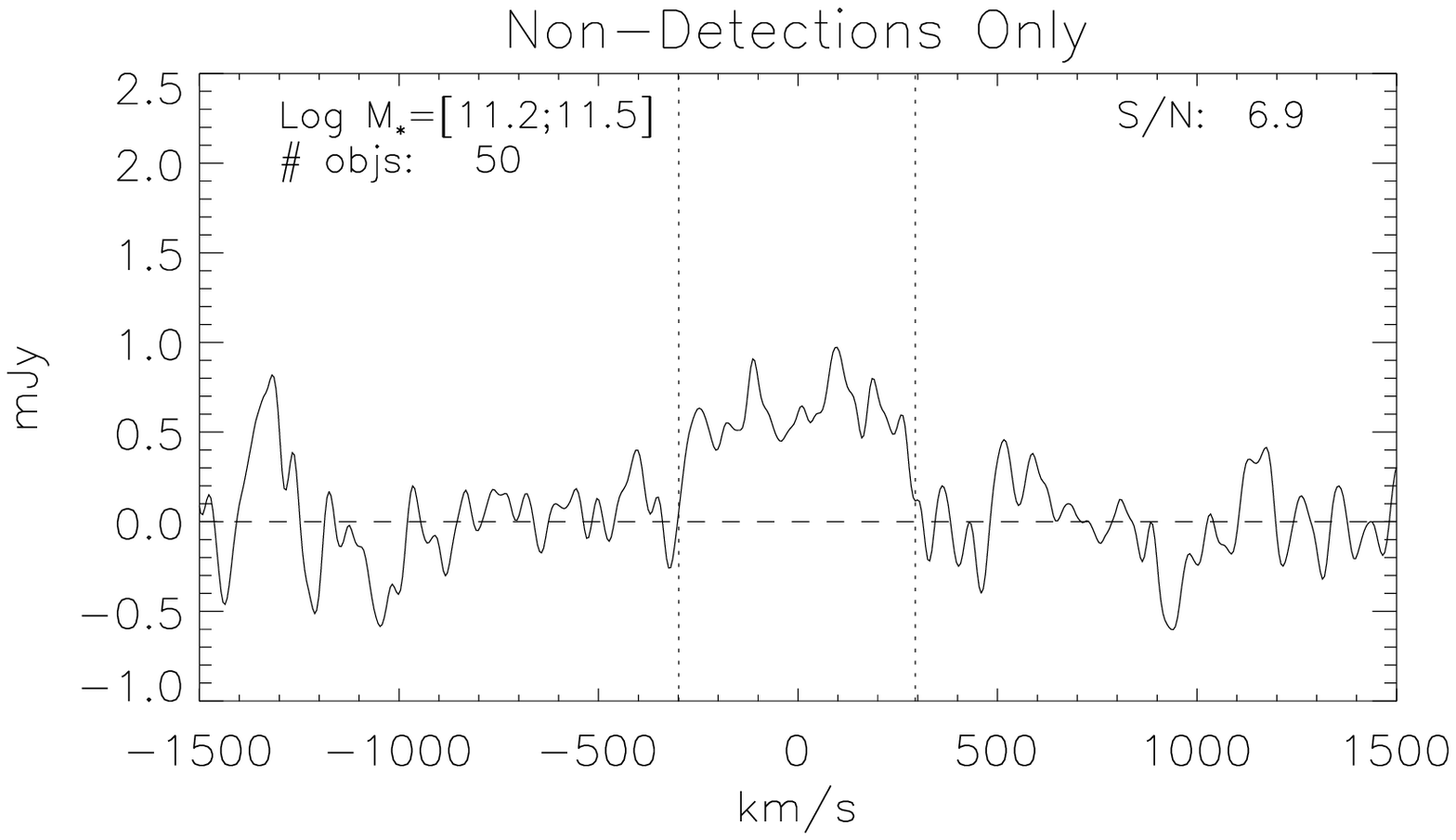}
\end {tabular}\caption{Examples of stacked spectra for 5 {\Mst} bins. The x-axis is velocity in km$\,$s$^{-1}$, the
  y-axis is {\hi} flux density in mJy. Since we shifted each object to its galaxy rest-frequency, the signal is
  centered at zero velocity. For each spectrum, the
  mass range and the number of objects stacked are reported, as well as
  its S/N ratio. Dotted lines show the boundaries of the
  signal, inside which we integrate the flux. In some spectra there are spikes/holes caused by
  poor quality data (note that the spectra containing the bad data were not discarded because the bad pixels are located away from the central regions of interest). Examples of bad regions occur 
  in the first two rows between $v\,=\,$-1500 km$\,$s$^{-1}$ and $v\,=\,$-1000 km$\,$s$^{-1}$.
  Left column: stacked spectra using all  galaxies.
  Right column: stacked spectra using only those galaxies that were not detected by ALFALFA.
  As expected, the signal is systematically lower for
  the right column.  }
\label{fig06}\end{figure*}

Figure \ref{fig06} shows some examples of stacked
spectra, obtained by stacking \textit{sample A} galaxies in five different 
stellar mass intervals. 
The vertical axis in each panel shows flux density in mJy, while the horizontal one
shows velocity in km$\,$s$^{-1}$. Since we shifted each
object to the rest-frequency of each galaxy, the stacked signal is centered at
zero velocity.  For each stacked spectrum, 
the mass range and the number of objects stacked are reported, as well as
the signal-to-noise ratio of the detected {\hi} signal. As explained above, when the
signal-to-noise is low, we smooth the spectrum over up to nine
channels as done, for example, for the last spectrum in Figure \ref{fig06}. 
 In the left column of the Figure, we show  spectra obtained
by stacking ALFALFA detections and non-detections. In the right panel
we show the spectra obtained by stacking only the
non-detections, demonstrating that the stacking process
recovers a signal even if no individual galaxy
is detected (dotted lines in Fig. \ref{fig06} indicate the edges of
the signal). Notice
that the \textit{rms} decreases with increasing number 
of co-added objects.  The width of the profile is
smaller for less massive objects, as expected since 
they have on average lower circular velocities.

If we recover a signal in the stacked spectrum, we measure the  
integrated  emission between the two edges of the {\hi} profile (see
dotted lines in Figure \ref{fig06}),
which are defined manually for each spectrum. 
We expect the signal to be symmetric around 
zero velocity. Even if in some cases the SDSS redshift is slightly
off-set with respect to the {\hi} emission, the discrepancy
will be random and will cancel out when co-adding multiple spectra 
(Note that the SDSS redshift
uncertainty is 0.0002, which corresponds to $\sim$60$\,$km$\,$s$^{-1}$).

We evaluate a signal-to-noise ratio following \citet{saintonge07}.
We define an {\hi}  detection if the  S/N is greater  than 6.5. 
If there is no detection,  we evaluate an upper limit, assuming a 5 $\sigma$
signal with a width of 300 km$\,$s$^{-1}$, smoothing the spectrum to 150
km$\,$s$^{-1}$. 
(The width of 300 km$\,$s$^{-1}$ is chosen because it corresponds
to the peak of the distribution of
velocity widths for galaxies in the GASS survey.)

\subsection{Evaluating {\hi} gas fractions}\label{stacktool_c}
Our aim in stacking  spectra is  to characterize the average {\hi}
 content of a given sample of
galaxies, so we are interested in converting our recovered 
 signal into an {\hi} mass and also into an average 
{\hi} gas fraction. For a single  object, these two quantities are well defined.
The {\hi} mass is evaluated using the \citet{roberts63} formula:
\begin{eqnarray}\label{eq:rob}
\frac{M_{HI}}{\mathrm{M_{\odot}}}=\frac{2.356\times10^5}{1+z}\left(\frac{D_L(z)}{\mathrm{Mpc}}\right)^2\left(\frac{S_{int}}{\mathrm{Jy~km~s^{-1}}}\right),
\end{eqnarray}
where $D_L(z)$ is the luminosity distance and $S_{int}$ the integrated
{\hi} line flux. A correction for {\hi} self-absorption is not applied to the {\hi}
mass, as it is likely to be negligible \citep[][Appendix B]{H&G84}. 
The {\hi} gas fraction is simply defined as $M_{HI}/M_*$. 

For N individual detections (each with a flux measurement), the   
\textit{average} value of their gas fractions can be defined  
by the weighted mean value:
\begin{eqnarray}
  \langle\frac{M_{HI}}{M_{*}}\rangle\,&=&\,\left(\Sigma_{i=0}^N\frac{M_{HI;i}}{M_{*;i}}\cdot
  w_{i} \right)/\left( \Sigma_{i=0}^N
    w_{i}\right) 
\end{eqnarray}
where the $w_i$ are defined as in  equation 3.2.  

When co-adding spectra, we need to take into account
the fact that our targets span a significant range in redshift
($z_{max}\,=\,2\cdot z_{min}$), so given the same {\hi} mass an object at the
lowest redshift limit contributes  4 times more signal than one at the
upper redshift limit. Because we are stacking mainly non-detected
spectra we do not know how much each galaxy  contributes to the total signal. 
Both the mean and median values of redshift  and stellar mass
may not be representative.  
In order to weight each spectrum in a consistent manner, we choose to stack 
``gas-fraction'' spectra $S'_i$, where
the signal that we co-add is no longer the flux $S\,$[mJy], but the quantity:
\begin{eqnarray}
S_i\,\mathrm{[mJy]}\,\rightarrow\,S'_i\,\mathrm{[mJy\,Mpc^2\,M^{-1}_{\odot}]}\,=\,\frac{S_{i}\cdot D^2_L(z_i)}{M_{*;i}}
\end{eqnarray} The stacking is performed according to equation
\ref{eq:stack}. With this approach, the stacking creates an average
object with respect to the {\hi} gas fraction of the individual
galaxies used to build the stack. 
In Appendix A, we discuss  an alternative approach to evaluating
mean gas fractions from stacked spectra, and present comparisons 
between the two different methods. 

We measure errors on the gas fractions using the \textit{Jackknife
  method} \citep{Tukey77, jack}, a statistical tool 
to estimate a confidence interval on a
function of N measures. The purpose of computing jackknife errors
on the  mean {\hi} fraction estimated from a stacked spectrum
is to ascertain whether or not the signal is dominated by a few outliers. 
Schematically, given a function $\hat{\alpha}(x_i)$ of $N$
observations $x_i$, the Jackknife evaluates $N$ partial estimates $\hat{\alpha}_j$ of the function obtained by discarding one
element per time. The \textit{jackknifed estimate} $\hat{\alpha}^*$ is then the average of the
pseudo-values $\hat{\alpha}^*_j=N\hat{\alpha}-(N-1)\hat{\alpha}_j$.
Finally, the sample variance on the pseudo-values is:
\begin{equation}\label{eq:jack}
    (\sigma_{Jack})^2 = \frac{1}{N(N-1)} \sum_{j=1}^N (\hat{\alpha}^*_j-\hat{\alpha}^*)^2,\\
\end{equation}
This can be used to provide  a confidence interval on $\hat{\alpha}-\hat{\alpha}^*$.

\section{{\hi} gas fraction scaling relations for massive galaxies}\label{stackGass}
\begin {figure*}
\begin {tabular}{c}
\includegraphics[width=14cm]{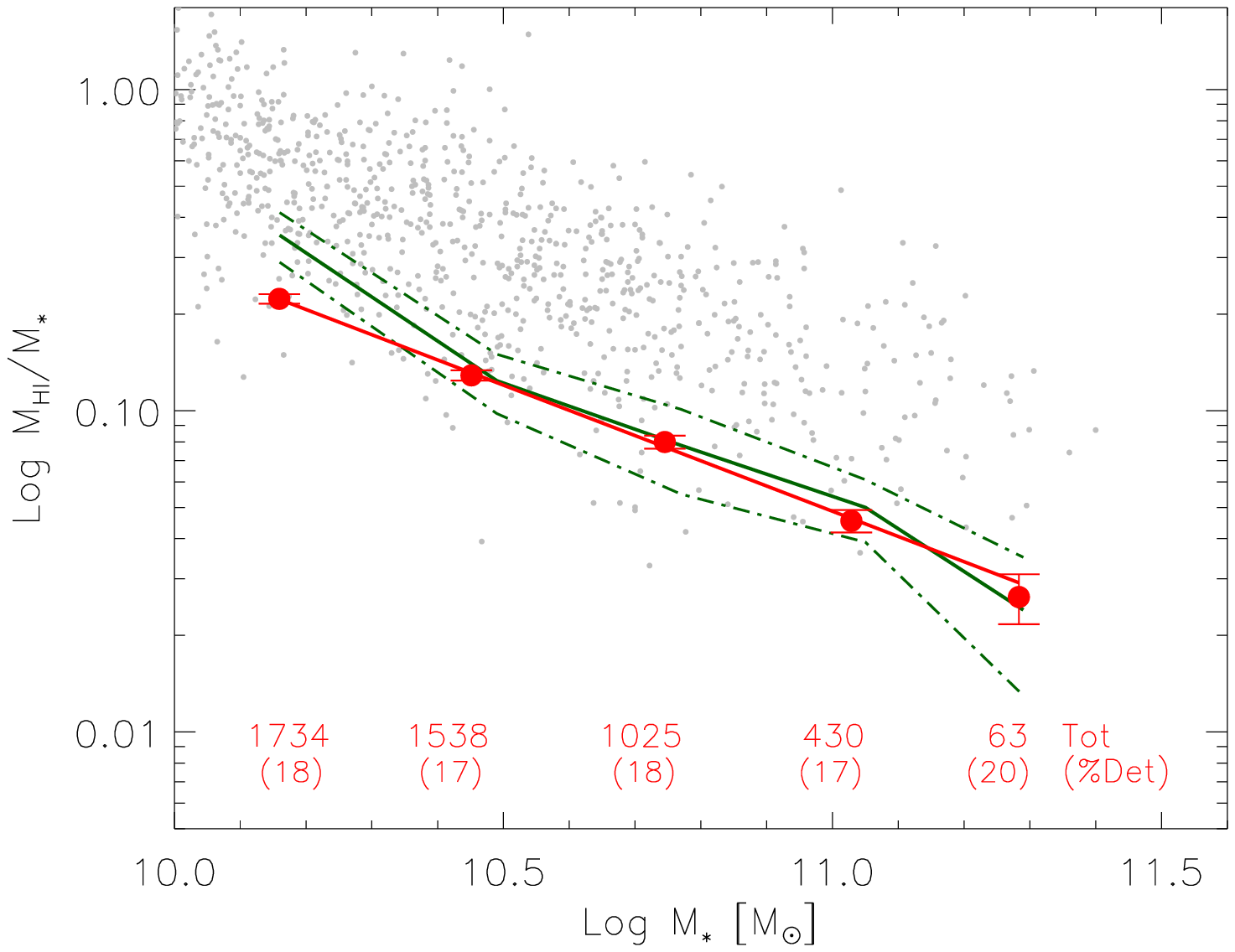}\\
\includegraphics[width=14cm]{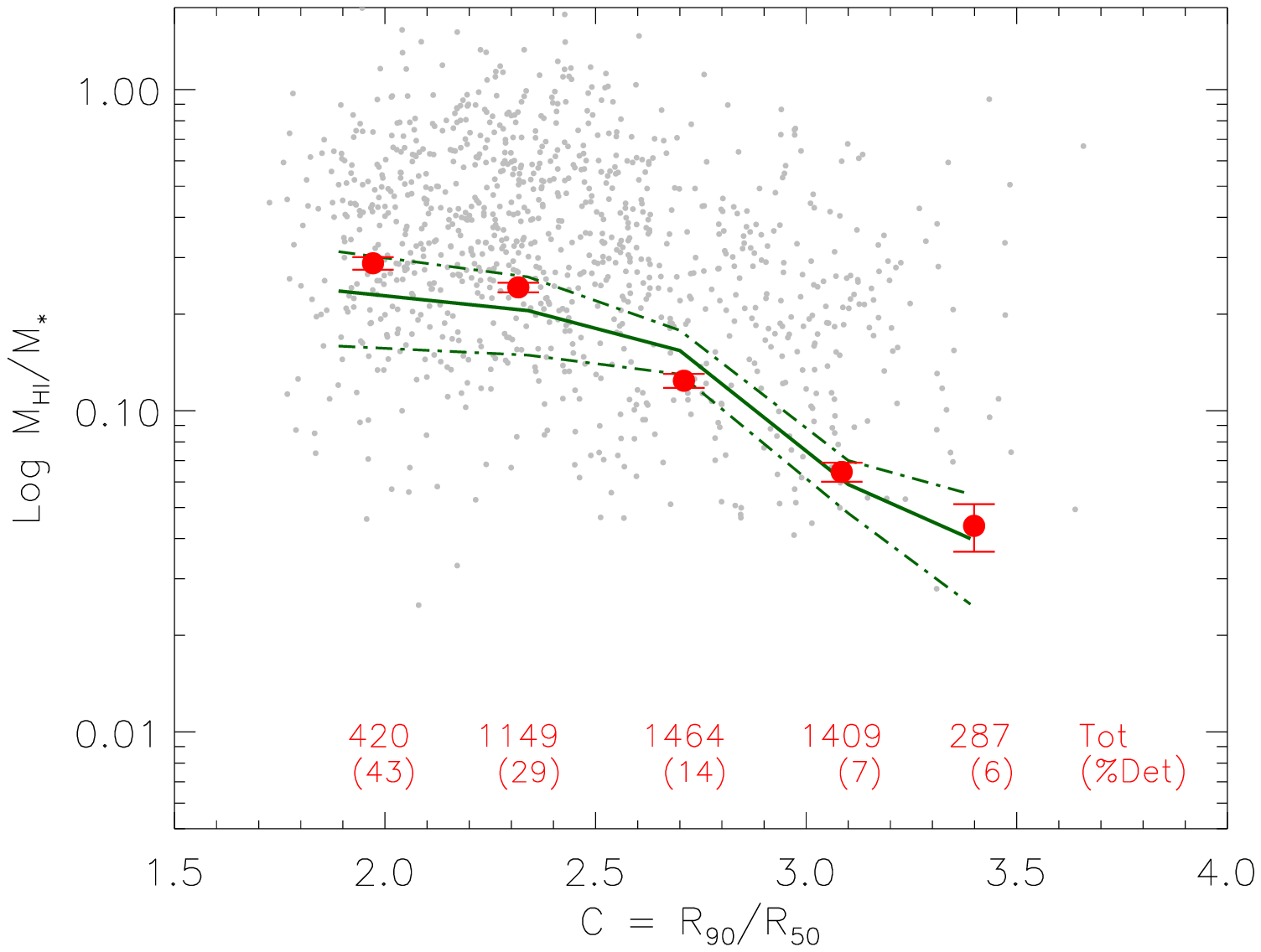}
\end {tabular}
\caption{Red circles: Average gas fractions of \textit{sample A}
  galaxies derived  from stacking are plotted as a function
  of stellar mass {$M_*$} (top) and concentration index $C$  (bottom). The points are plotted at the mean
 value of  $M_*$ or $C$  for the galaxies in the bin. The red line
 (upper panel) show a linear fit to these results, which are compared
 with the average {\hi} gas fractions from the GASS-1 paper
  (green lines). Green dashed lines show the 
 1$\sigma$ uncertainties on the GASS-1 estimates. Gray dots show
 galaxies with ALFALFA detections from \textit{sample A}. The numbers
 written in the panels indicate the numbers of objects co-added in
 each bin (Tot), and the percentage of them directly detected by
 ALFALFA (\%Det). The values plotted are reported in Table
 \ref{tab01}, 4th and 5th columns.}\label{fig07}
\end{figure*}
\begin {figure*}
\begin {tabular}{c}
\includegraphics[width=14cm]{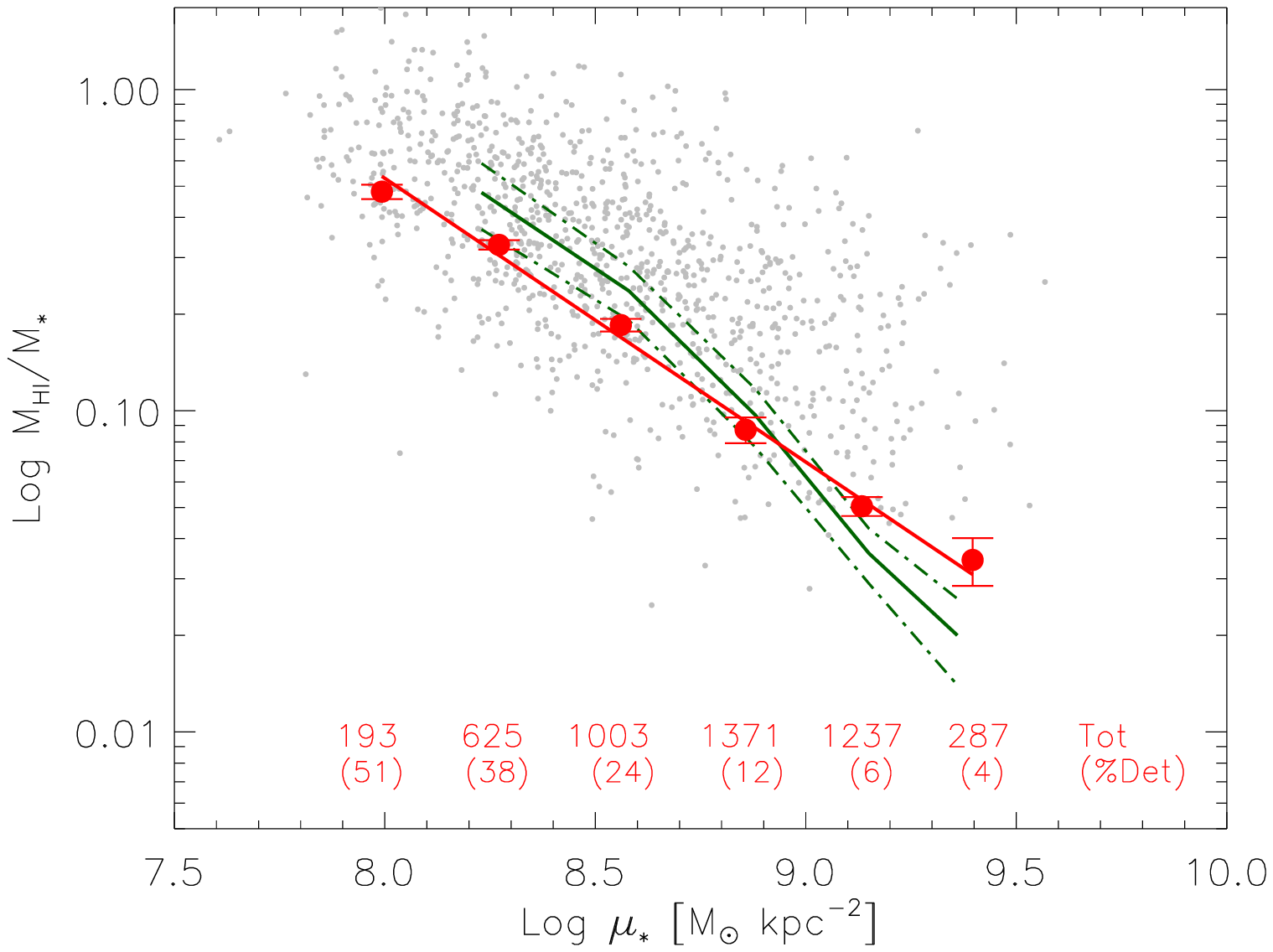}\\
\includegraphics[width=14cm]{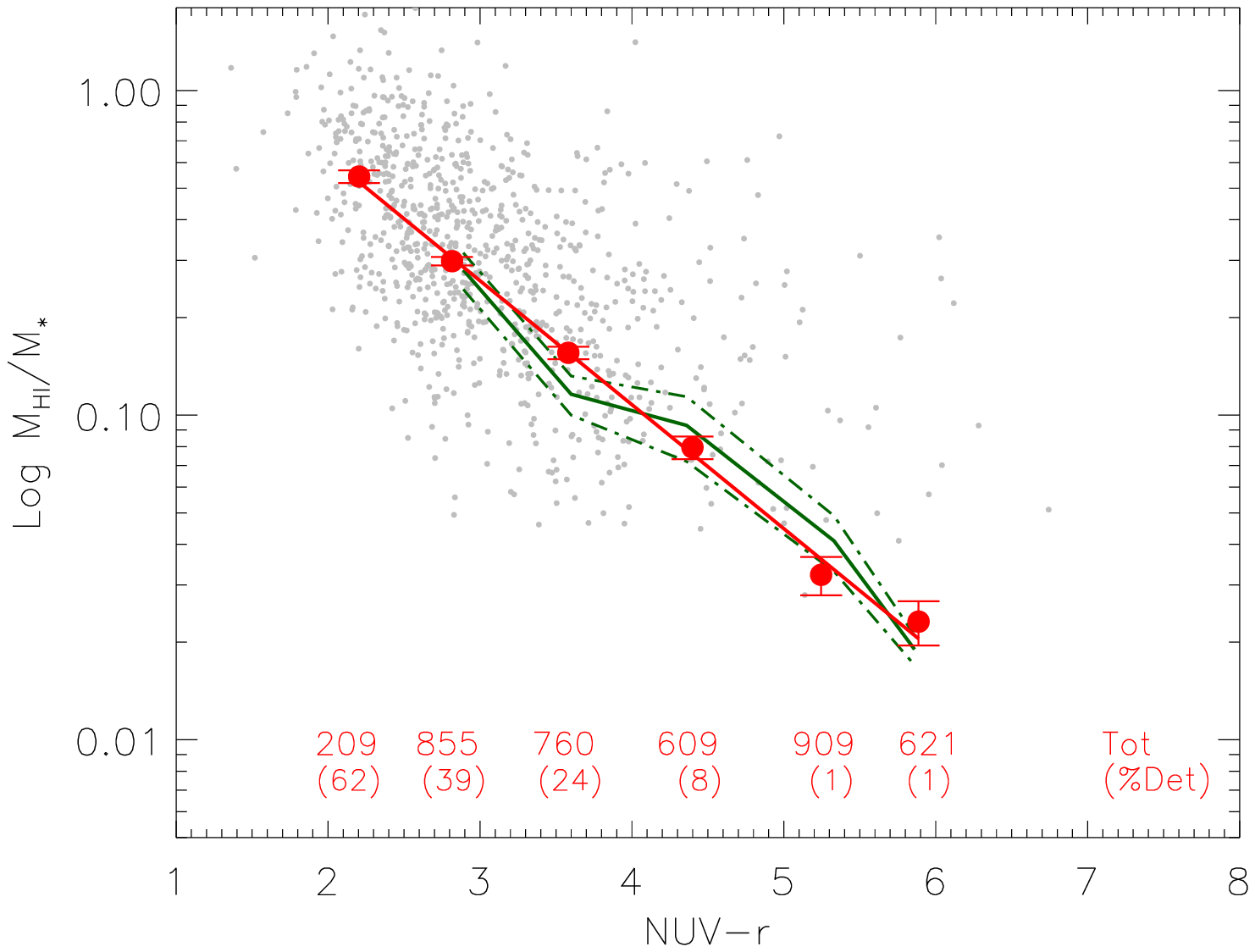}
\end {tabular}
\caption{Average gas fractions of \textit {sample A} galaxies derived from stacking are plotted as a function
  of stellar mass surface density {\must} (top) and {\col} colour
  (bottom). Symbols and colours as in Figure \ref{fig07}. The values plotted are reported in Table
 \ref{tab01}, 4th and 5th columns.}
  \label{fig08}
\end{figure*}
In this section, we characterize how average {\hi} gas fraction
depends on a variety of different properties for galaxies 
with stellar masses greater than $10^{10} M_{\odot}$. First we study
the dependence on stellar mass {\Mst}, stellar mass surface
density {\must}, concentration index {\cix} and
colour {\col} for galaxies in  \textit{sample A}. We then compare the same relations for galaxies in the 
bulge-dominated \textit{ETG sample} ($\S$ \ref{etg}).

We begin by  comparing our results with
those of  Catinella et al. (2010), who  studied the correlation between {\hi}  gas
fraction and the  properties listed above, for a  complete  sample of
$\sim$ 200 galaxies from the GASS survey  with $M_* > 10^{10} M_{\odot}$. In that survey, 
the {\hi} observations reached 
a much deeper flux limit  compared with  ALFALFA, so   
average {\hi} gas fractions could be derived using individual {\hi} detections, rather
than by stacking. It is important to check whether our stacking method
gives answers that are consistent with these published measurements.
 
In Figures 7 and 8, we present our results for  \textit{sample A}.
In each panel, gray dots show the galaxies in  \textit{sample A} that were detected by ALFALFA.
 Green lines show the average values and 1$\sigma$ confidence
interval of {\Mhi}/$M_*$ from          
GASS-1 (see Figure 9 in that paper). Our own estimates of
the {\hi} gas fraction derived from the stacked spectra 
are plotted as red circles, and the errorbars on these points are evaluated using the  
\textit{Jackknife method}
(eq. \ref{eq:jack}). Note that the stacked spectra that yield the
measurements plotted as red circles in Figure \ref{fig07} (top panel),
are shown in Figure
\ref{fig06} (left panel). At the bottom of each panel, we record  the number of galaxies
included in each stack, as well as the percentage of galaxies
with ALFALFA detections. In cases
where the stack includes more than a few hundred galaxies,
the errors on the mean {\hi} fraction are  negligible. 

We note that we are able to measure an {\hi} gas fraction in
every bin, even in  the very reddest NUV-r colour bins
where the ALFALFA detection rates are close to zero. The fact that the jackknife errors
remain relatively small for these bins, indicates that the   
stacked spectra are not dominated by signal from a small fraction of
the galaxies. \\

We find that the average {\hi} gas fractions derived from the stacked spectra are in excellent
agreement with the results reported in GASS-1.  The point-by-point agreement is generally
within 1$\sigma$, with the possible exception of the lowest stellar
mass and stellar mass surface density bins, where we find gas fractions that are systematically
lower than those reported in the GASS-1 paper. We caution that low mass galaxies were somewhat 
under-represented in the first GASS data release. Our errorbars 
only indicate statistical errors and do not account for other effects, such as
cosmic variance, so we do not think the discrepancy is significant. 

As discussed in GASS-1, the {\hi} gas fraction is a decreasing function
of stellar mass, stellar mass surface density, concentration index and colour.
The slopes of these relations for our \textit{sample A} are
  listed in the 3rd column of Table \ref{tab01}. Here we only report
  the total change in $\log M_{\rm HI}/M_*$ ($\Delta \log M_{\rm
    HI}/M_*$) for each relationship, which allows a comparison among
  the different properties:
  \begin{eqnarray*}
    M_*&:\qquad \Delta \log M_{\rm
      HI}/M_*\,=\,0.20\\
    C&:\qquad \Delta \log M_{\rm
      HI}/M_*\,=\,0.24\\
    \mu_*&:\qquad \Delta \log M_{\rm
      HI}/M_*\,=\,0.45\\
   \rm NUV-\textit{r}&:\qquad \Delta \log M_{\rm
      HI}/M_*\,=\,0.52
  \end{eqnarray*}
The correlations between gas fraction and both {\Mst} and concentration index
(Figure \ref{fig07}) are weak. In contrast, the average value of $M_{\rm HI}/M_*$ drops by
more up to a factor of 25 when plotted  
as a function of NUV-r colour or stellar mass surface density.
Although stacking  recovers the average trends of the
population extremely well, it does not provide any information on the
underlying  scatter. 
This  can only be studied  with individual gas fraction
measurements, as in GASS-1.

\section{{\hi} study of a complete sample of ETG galaxies}\label{etg}
\begin {figure*} 
\begin {tabular}{c}
  \includegraphics[width=14cm]{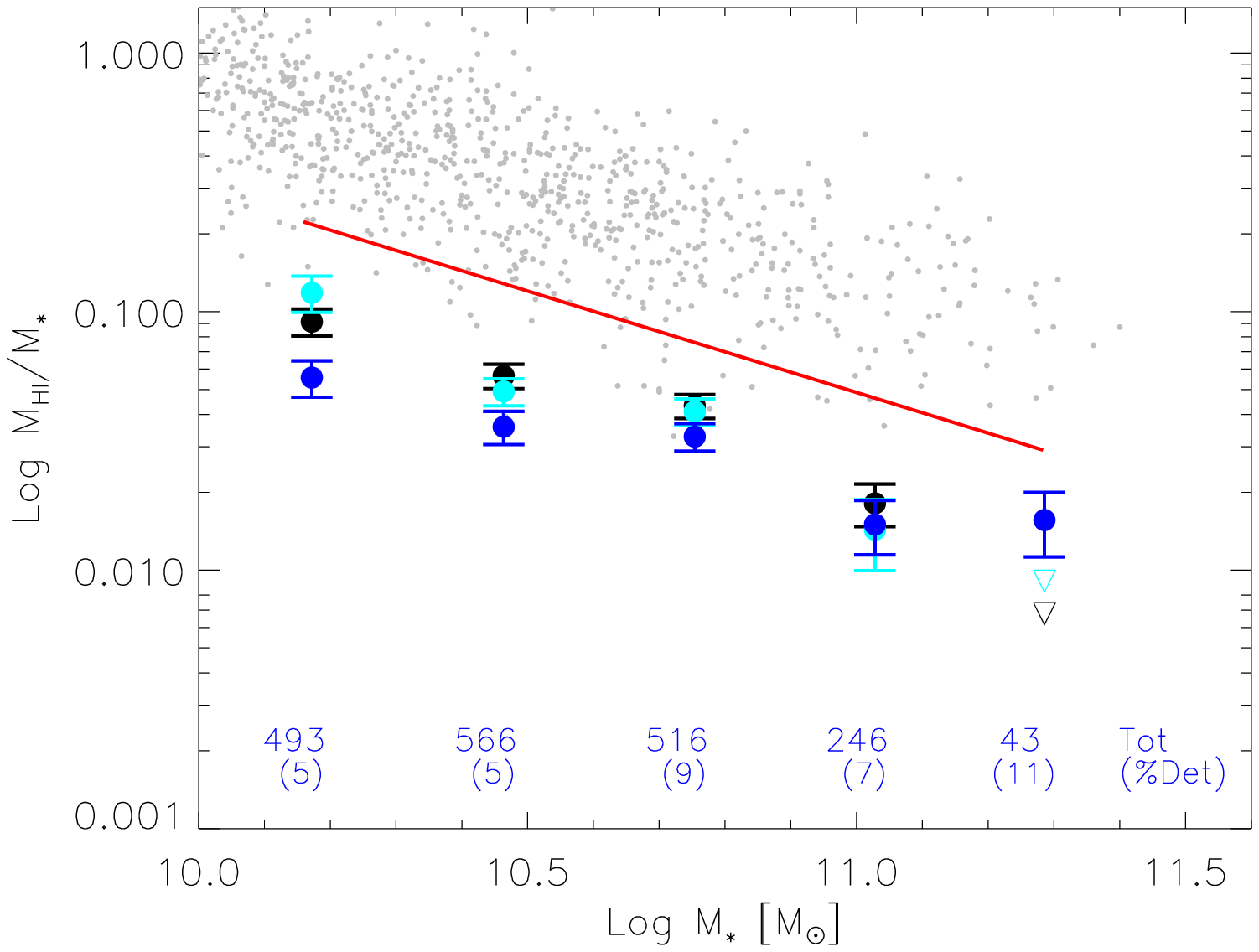}\\
\includegraphics[width=14cm]{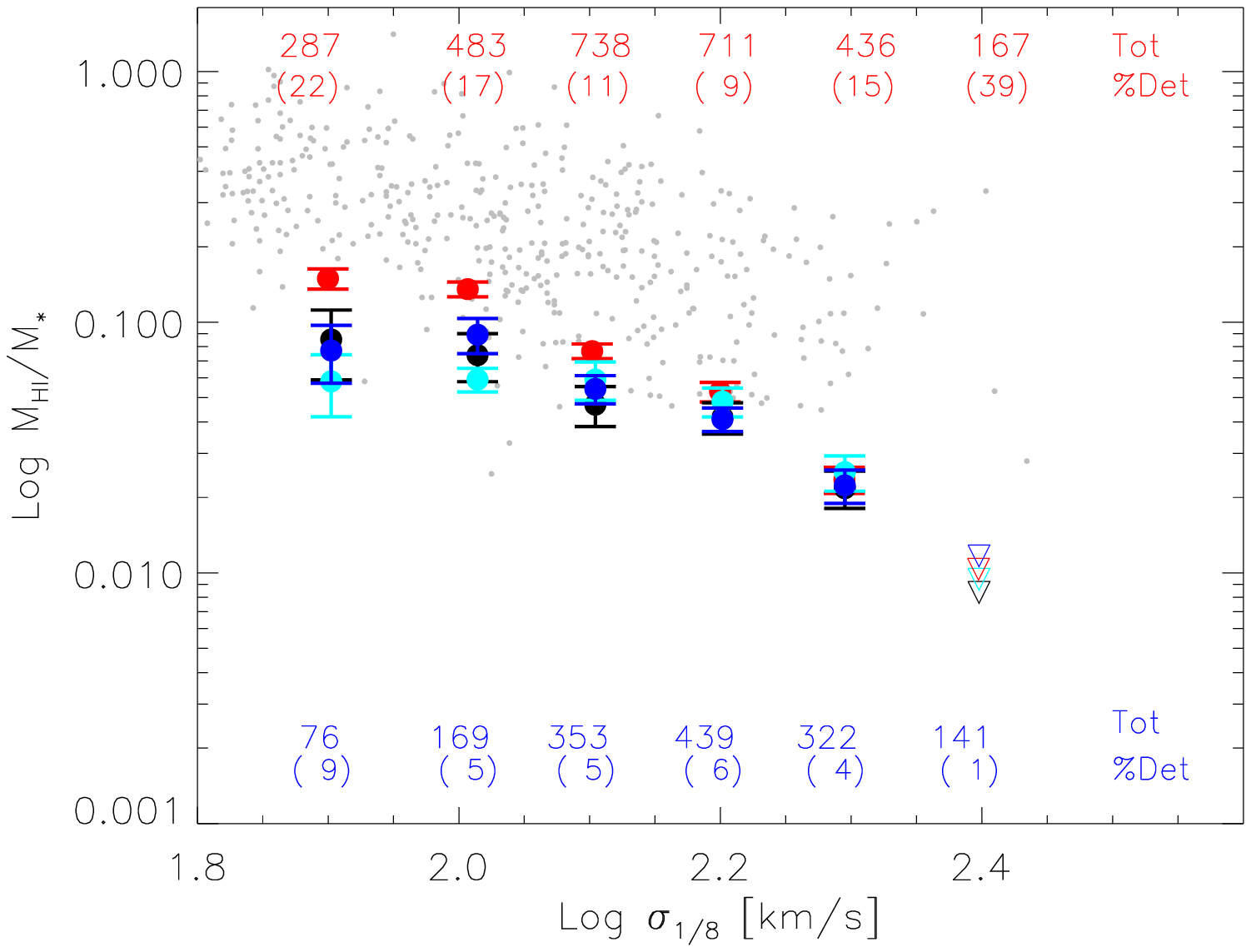}
\end {tabular}\caption{ The dependence of the average {\hi} gas 
fraction on stellar mass {\Mst} (top)
and on central velocity dispersion $\sigma$ (bottom) for  the
\textit{ETG sample} (blue symbols). The relations found for
\textit{sample A} are shown in red for comparison (the fit for the
       {\Mst} relation, the actual points for $\sigma$). 
Upside-down triangles indicate upper limits in the case of a non-detection.
 The numbers
 written in the panels indicate the numbers of objects co-added in
 each bin (Tot), and the percentage of them directly detected by ALFALFA (\%Det). 
Gray dots show \textit{sample A} galaxies with ALFALFA detections. We have also
applied more stringent cuts to the \textit{ETG sample}, as explained  in the
text: cyan circles represent a sample with  {\cix}$>$3
(52\% of the original ETG sample); black circles are for a sample with
b/a$\, >$0.6 (or inclination lower than 55$^{\circ}$ - 68\%
of the original sample). The values plotted for each \textit{ETG
  sample} are reported in Table
 \ref{tab01}, 6th-11th columns.}
\label{fig09}
\end{figure*} 
\begin {figure*} 
\begin {tabular}{c}
\includegraphics[width=14cm]{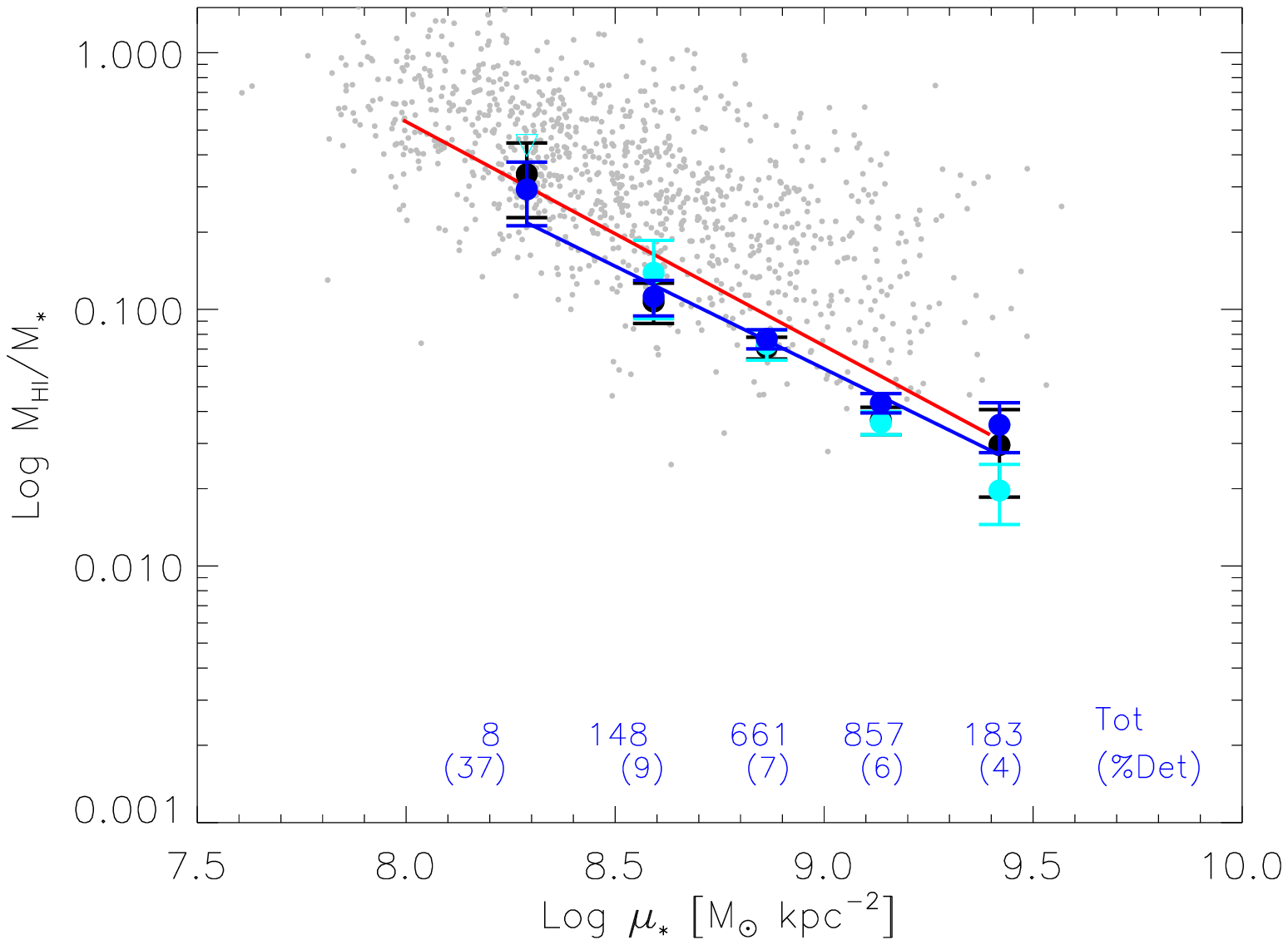}\\
\includegraphics[width=14cm]{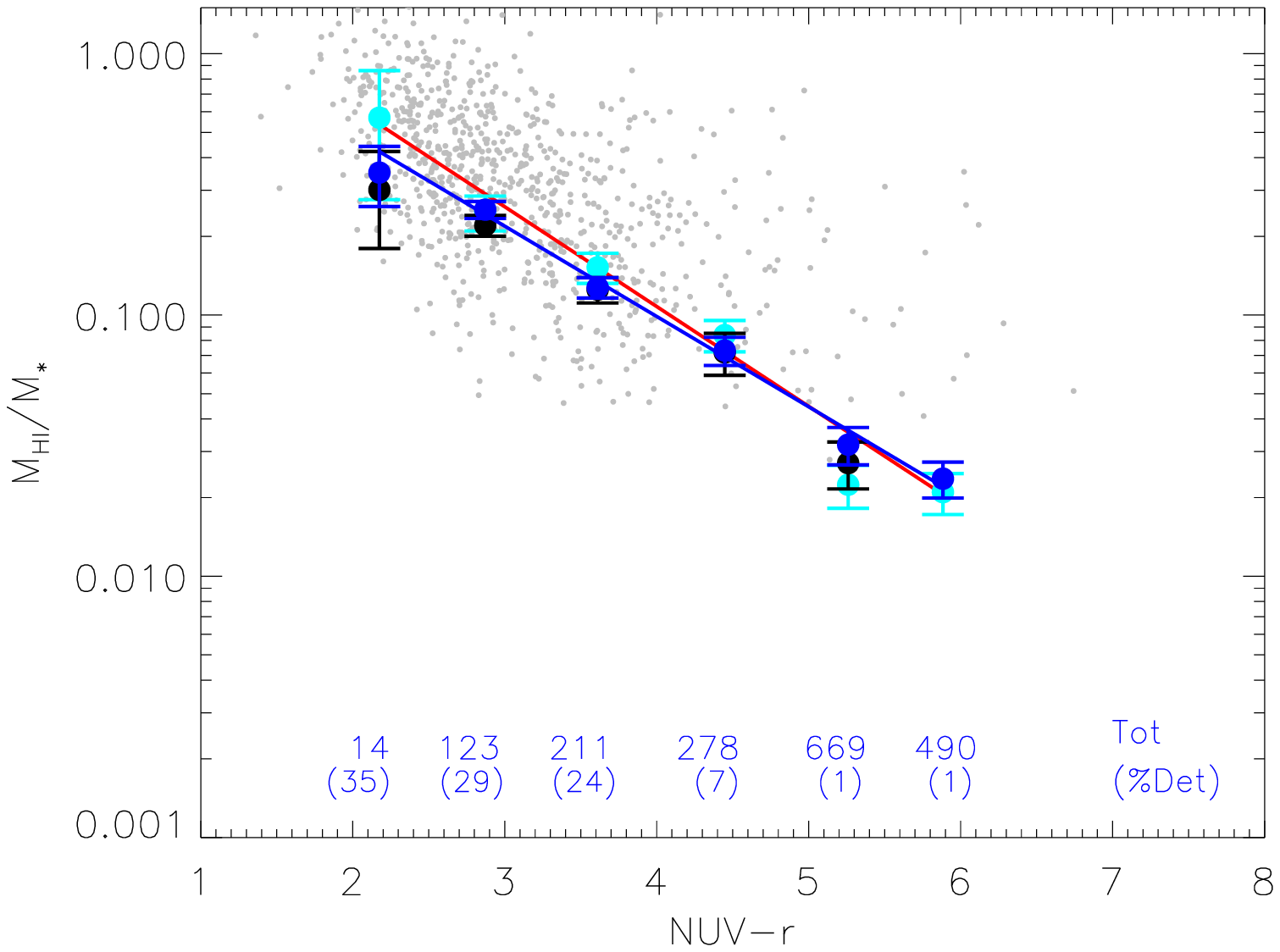}
\end {tabular}\caption{Average {\hi} gas fraction dependence on stellar mass surface
  density {\must} (top) and {\col} colour (bottom) for  the
  \textit{ETG sample}. Symbols and colours are the same as described
  in Figure \ref{fig09}. The values plotted for each \textit{ETG sample} are reported in Table
 \ref{tab01}, 6th-11th columns.}
\label{fig10}
\end{figure*} 
\begin {figure*} 
\begin {tabular}{c}
\includegraphics[width=8cm]{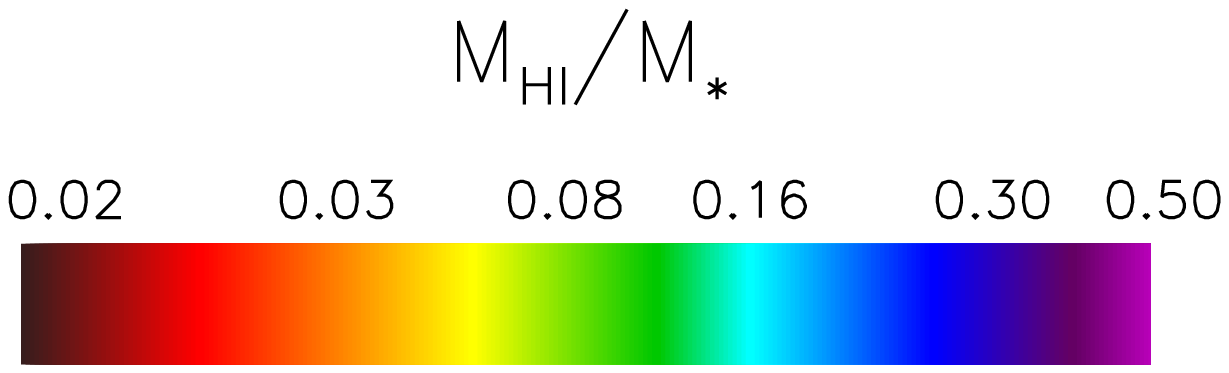}\\
\includegraphics[width=12.8cm]{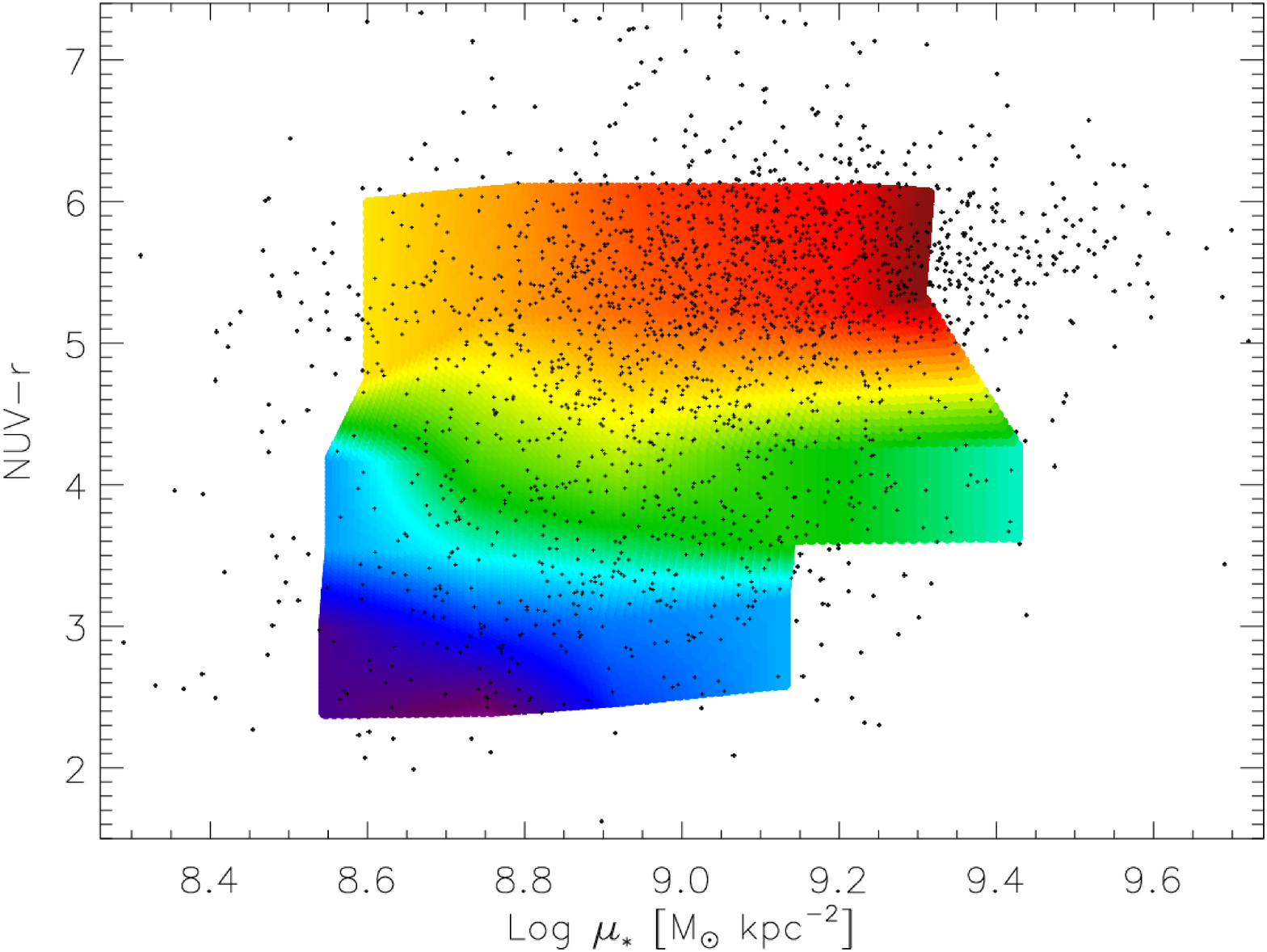}\\
\includegraphics[width=12.8cm]{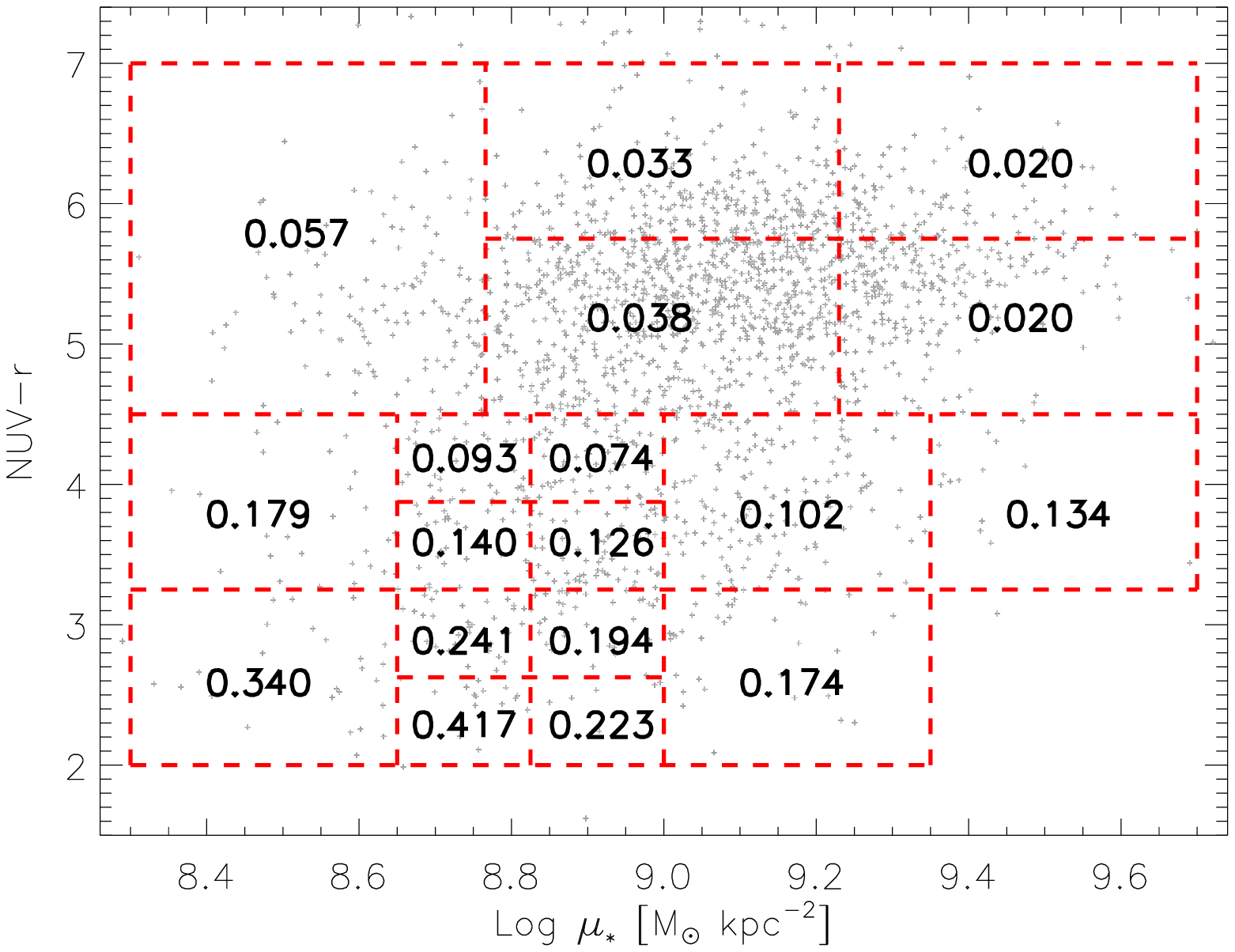}
\end {tabular}\caption{Average {\hi} gas fraction dependence in the 2-dimensional plane of 
stellar mass surface density {\must} and {\col} colour for galaxies in  the
  \textit{ETG sample}. In the top panel the dots show individual
  objects, while the colors show the (interpolated) gas fractions
  measured with the stacking as a function of position in the
  plane (the colour scale key is included above the plot). The bottom
  panel shows the adopted binning. For each
  bin,  the gas fraction measured from the stacked spectrum is noted.}  
\label{fig11}
\end{figure*} 
In this section, we ask whether bulge-dominated galaxies
with  $M_* > 10^{10} M_{\odot}$ lie on the same {\hi}
scaling relations as the general population of galaxies with $M_* > 10^{10} M_{\odot}$.
Our goal is to determine whether the presence of the bulge plays a role 
in regulating the rate at which gas is consumed into stars, 
for example by stabilizing the disk \citep[e.g.,][]{Martig09}. In
order to test the role of the bulge in a clean way, we must account for
the fact that the  physical  properties of galaxies are strongly correlated. 
If one selects a sub-sample of bulge-dominated galaxies from
the parent \textit{sample A}, one will automatically select a sample of galaxies
with higher stellar masses, higher stellar mass surface densities, and redder colours.
It is therefore important to understand whether or not bulge-dominated galaxies
differ in {\hi} content from the parent sample at {\it fixed} values of these
parameters.
 
Our results are presented in  Figures \ref{fig09} and
\ref{fig10} and in Table \ref{tab01}. In all the plots, blue circles are the average
gas fractions obtained from stacking the \textit{ETG sample} and the errorbars are evaluated using
the \textit{Jackknife method} (eq. \ref{eq:jack}). Upside-down triangles indicate the upper limit in
the case of a stack that yields a  non-detection.
The red lines (circles) show the fits to the mean {\hi} gas fraction relations obtained for
\textit{sample A}. Our main result is that the average {\hi} gas fractions of  bulge-dominated galaxies
are significantly lower (by approximately a factor of 2) than those of the parent sample at a given value of stellar mass.
A similar, but weaker reduction in the average {\hi} gas fraction is seen for the \textit{ETG sample} when 
it is plotted as a function of the central velocity dispersion of the bulge.
However, the relation between  gas fraction and 
stellar mass surface density and  NUV-r colour appears to be insensitive to the ETG cut.    
In  GASS-1 paper, Catinella et al. showed that a linear combination of stellar mass surface density and NUV-r
colour provided an excellent way to ``predict'' the {\hi} content of galaxies more massive than
$10^{10} M_{\odot}$. Here we show that this conclusion holds
true independent of the bulge-to-disk ratio of the galaxy. 
We have tested that this conclusion still holds if we define the \textit{ETG sample} using more
stringent cuts on concentration index ($C>3$) or on the axis ratio of the
galaxy ($b/a>0.6$, which implies inclination lower than
$\sim$55$^{\circ}$). These cuts reduce the ETG by  50\% and 30\%, respectively.  
Nevertheless, results shown in Figures 9 and 10 demonstrate that the average {\hi} gas fractions
of these systems still lie on the same relations when plotted as a function of
stellar mass surface density and NUV-$r$ colour (cyan circles
represent the further cut in concentration index, black ones the cut in inclination).

In Figure \ref{fig11}, we show how the average gas fractions of 
galaxies in the \textit{ETG sample} vary as a function of position in the
two-dimensional plane of colour versus stellar mass density $\mu_*$.
Bulge-dominated galaxies are mainly found on the red sequence,
but there is a minority population with bluer colours. 
We adaptively bin the sample in two dimensions  
by recursively dividing the plane into
axis-aligned rectangles. We stop dividing a region  when a further split
would lower the S/N below the detection threshold of 6.5. 
Figure \ref{fig11} (bottom panel) shows the final binning used. In
each bin the measured gas fraction (expressed as a percentage
of the stellar mass) is reported. 
In Figure \ref{fig11} (top panel) we colour-code the (NUV$-r$)$-\mu_*$ plane
according to gas fraction. 
The {\hi} content decreases going from left to right (towards increasing
stellar mass surface density) and from bottom to top (towards redder
colours).  The most significant variation is clearly along the colour direction.

\subsection{A Test of the morphological quenching scenario}\label{MQ}
\begin {figure*} 
  \includegraphics[width=12.8cm]{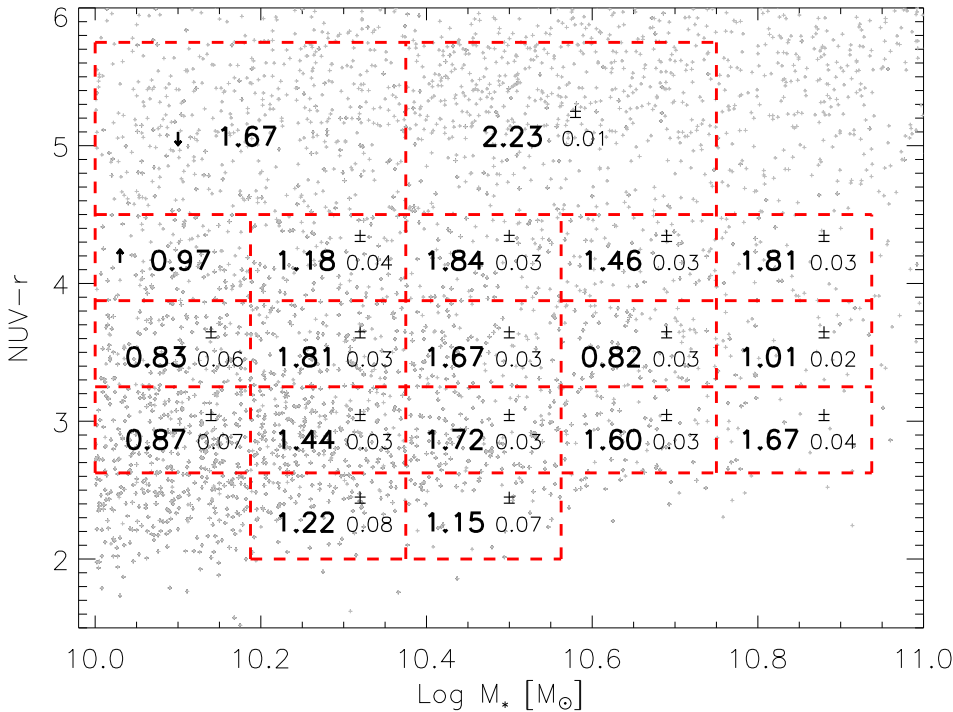}
\caption{Morphological quenching test. For each bin of
  {\col} and {\Mst} the ratio between the gas fractions of the
  disk-dominated galaxies with $C<2.6$ f(HI)$_{DD}$ and the bulge-dominated
  galaxies with $C>2.6$ f(HI)$_{BD}$ is reported. 
 The panel shows the adopted binning with the values measured for each
  bin. The arrows indicate upper/lower limit if one of the stacked
  spectra yielded a non-detection. Gray dots show how the targets
  (LTGs+ETGs) distribute along the plane.}\label{fig12}
\end{figure*} 
The idea that galaxy disks are more resistant to the formation of bars, spiral
density waves and other instabilities if they are embedded within a 
dynamically hot halo or bulge, has its origins in early work by \citet{O&P73}. Recently, \citet{Martig09} have proposed   
this so-called  ``morphological quenching'' mechanism as a way of explaining  
why present-day bulge-dominated galaxies on the red sequence cease growing  in
stellar mass in environments where they continue 
to accrete gas. In their picture, a disk
with similar gas content will be much less efficient at forming stars if it
is embedded in a galaxy with a significant 
bulge component. As stated in the abstract of their
paper, ``our mechanism automatically links the colour of the galaxy with its
morphology, and does not require gas consumption, removal or termination 
of the gas supply.''

To test whether the ``morphological quenching'' process is truly important in
maintaining the low observed rates of star formation in red sequence
galaxies, we have performed the following experiment.
We have binned galaxies with $C<2.6$ and $C>2.6$ in the two-dimensional
plane of NUV-$r$ colour versus stellar mass (note that we use the same
bin boundaries for both samples). We stack the {\hi} spectra of the  galaxies  
in each bin and calculate the average {\hi} gas fraction, as explained above.
In Figure \ref{fig12},  we report for each bin  the ratio
between the gas fraction of the disk-dominated (\emph{DD}) objects and the
bulge-dominated (\emph{BD}) ones, i.e.: 
\[r\,=\, \left(\frac{M_{HI}}{M_*}\right)_{DD}/\left(\frac{M_{HI}}{M_*}\right)_{BD}\]
If the morphological quenching scenario is correct, then at fixed
stellar mass,  we would expect to
find  higher average {\hi} gas fractions for bulge-dominated galaxies
on the red sequence than for disk-dominated galaxies on the red sequence.

The results in Figure \ref{fig12} show that in general the {\em opposite} is true.
Gas fractions are always slightly higher for the disk-dominated
galaxies than for bulge-dominated ones. The gas fraction differences
do appear to be largest for red sequence galaxies with NUV-$r$ $> 5$, but
the sign of the difference  contradicts the predictions of \citet{Martig09}. 

It is important to check that our result is not simply
due to extinction effects. Some of the  reddest, gas-rich disk-dominated
objects  may actually be heavily obscured systems
that will move blueward when dust corrections are applied.
 Following \citet{gass2} we have applied dust corrections to  the NUV-r
colours of the  galaxies in our sample with
D$_n$(4000)$<$1.7\footnote{ D$_n$(4000) is defined as the ratio of the average flux density in the
  continuum bands 3850-3950 and 4000-4100 {\AA}, and
  traces the age of stellar populations \citep{Kau03b}.}.
In order to make up for the loss of red late-type galaxies, we had to
apply stronger cut in concentration index 
(C$>$3) to define the early-type sample. 
The ratio of gas fraction between disk-dominated and
bulge-dominated objects decreases, but we still
find that the {\hi} gas fraction of the disk-dominated galaxies
\textit{never} falls below that of the bulge-dominated ones. In
particular, if we divide the red sequence 
objects with 4.5$\,<\,${\col}$\,<\,$7 in
two bins of stellar mass, we find a value of $r$=1.03 for the less
massive galaxies, and $r$=1.71 for the more
massive ones.     

We note that these results are in agreement with those presented in \citet{gass2}.
In their paper, Schiminovich et al. use a volume-limited sample of 200 galaxies
from the GASS survey to explore the global scaling relations associated
with the ratio SFR/{\Mhi}, which they call the {\em {\hi}-based star formation efficiency}.
They found that the average value of this star formation efficiency has little
variation with any galaxy parameter, including the concentration index.

\section{Summary}
We have carried out a stacking analysis using ALFALFA scans of a volume-limited sample of 
$\sim$ 5000 galaxies with imaging and spectroscopic data from  GALEX and the Sloan Digital Sky Survey. 
The galaxies have  stellar masses greater than $10^{10} M_{\odot}$ and redshifts 
in the range $0.025<z<0.05$. We extract a sub-sample of 1833 ``early-type'' galaxies
with inclinations less than 70$^{\circ}$, with concentration indices $C>2.6$, and with light profiles
that are well fit by a De Vaucouleurs model. We then stack the ALFALFA spectra of the
galaxies from these two samples in bins of stellar mass, stellar
surface mass density, 
central velocity dispersion, and NUV-$r$ colour, and we use the stacked spectra 
to estimate the average {\hi} gas fractions {\Mhi}$/M_*$ of the galaxies in each bin.

Our main result is that the {\hi} gas fractions of both early-type and late-type galaxies
correlate {\em primarily}  with  NUV-$r$ colour and stellar mass surface density. The relation between
average {\hi} gas fraction and these two parameters is {\em independent} of $C$, and hence
of the bulge-to-disk ratio of the galaxy. We note that at fixed stellar mass, early-type galaxies
  do have lower average {\hi} fractions than late-type galaxies, but this effect does not
arise as a direct consequence of the presence of the bulge and we
discuss possible implications below.

We have also tested whether the average {\hi} gas content
of bulge-dominated  galaxies differs from that of
late-type galaxies at fixed values of {\col} and {\must}. 
We find no evidence that red-sequence galaxies with a significant bulge component are less
efficient at turning their available gas reservoirs
into stars. This result is in contradiction with the ``morphological
quenching'' scenario proposed by \citet{Martig09}. 

\section{Discussion}
We now consider possible implications of this work.

{\bf 1. The {\hi} content of a galaxy is independent of its bulge-to-disk ratio}.
This can be understood if the following two conditions are
satisfied: 1) The {\hi} gas in early-type galaxies is always associated with
disks, or with material that is in the process of settling into disks.
2) The formation of the galactic disk is decoupled from the formation
of the bulge.  One  way that condition (2) could be satisfied is if the
disk is formed from gas that accretes {\em well after the bulge formation
event}, so that the amount and the configuration of the accreted material
is not in any way related to  properties of the bulge, such as its mass
or its velocity dispersion.

We now attempt  to establish whether this hypothesis is correct.
At fixed {\em total stellar mass}, we found that galaxies with
larger {\cix} have lower gas fractions than galaxies with
smaller {\cix} (Figure \ref{fig09}, upper
panel). If the gas is associated mainly with the disk, then
the ratio {\Mhi}/$M_{*;d\!i\!s\!k}$ should \emph{not} depend on {\cix}.
We have fit the \citet{Wein08} relation between B/T and {\cix}. Our bi-linear fit result is
$C=1.19+2.28\cdot B/T$. For each galaxy in our sample,
we compute  $M_{*;d\!i\!s\!k}$.  
In  Figure 13, we plot {\Mhi}/$M_{*;d\!i\!s\!k}$ as a function
of $\log M_*$ for \emph{ETG} and for  \emph{sample A} galaxies.   
As can be seen, the difference in gas fraction between the two samples is greatly
reduced when the HI mass is divided by the disk stellar mass. 
For reference, the red and blue
lines on the plot show {\Mhi}/$M_{*;t\!o\!t\!a\!l}$, 
as in Figures \ref{fig07} and \ref{fig09} (upper panels).

We note that our result that the {\hi} fraction of  early-type galaxies
depends on the size of the galaxy, but not on its bulge properties, is
opposite to what is found for bulge  stellar populations.  In a recent
paper, \citet{Graves09b} carried out an analysis of 16,000 nearby
quiescent galaxies with 3-arcsecond aperture spectra from the Sloan
Digital Sky Survey. Their paper demonstrates convincingly that mean
stellar age, [Fe/H], [Mg/H], and [Mg/Fe] scale strongly with the velocity
dispersion of the bulge, but there is  no  dependence on $R_e$ at a fixed
value of  $\sigma$. We thus infer that the  star formation {\em history}
of the central bulge does not depend on the  size of the galaxy. Our own
results show that the present-day gas content of early-types does depend
on size. This again argues for  bulge and disk formation processes that
are decoupled.
\begin {figure} 
\includegraphics[width=8.5cm]{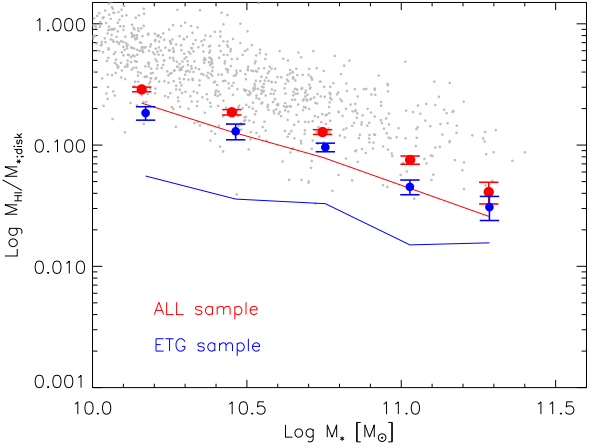}
\caption{Average \emph{disk} gas fractions of \textit{sample A} (red
  circles) and \textit{ETG sample} galaxies (blue circles) as a function
  of total stellar mass {$M_*$}. The lines show the results obtained
  measuring the total gas fraction, as in Figures \ref{fig07} and \ref{fig09}
 (upper panels). Gray dots show
 galaxies with ALFALFA detections from \textit{sample A}.}
\label{disc}
\end{figure} 

{\bf 2. Galaxies with significant bulge component are not less efficient
in turning their available gas into stars}. One possible explanation of
this result is that the rate of gas consumption in galactic disks is
primarily regulated by externally-driven rather than internally-generated
instabilities. In recent work, \citet{C&B09} present
an analysis of the observed perturbations of the {\hi} disk of the
Milky Way and infer the existence of a dark sub-halo that tidally
interacted with the Milky Way disk. In addition to dark sub-halos,
luminous satellite galaxies are observed to interact with
galactic disks.  Finally, the dark matter environment at the centers of
present day halos is neither static nor in equilibrium. \citet{G&W06}
used the Millennium Simulation to study asymmetries in dark
matter halo cores. They found that 20 per cent of cluster haloes have
density center separated from barycenter by more than 20 percent of the
virial radius, while only 7 percent of Milky Way haloes have such large
asymmetries. Because early-type galaxies reside in more massive dark
matter halos than late-type galaxies \citep{Mandel06}, their disks
may be subject to considerably larger  externally-driven perturbations.
All these effects may counteract the stabilizing effect of the bulge.

Further tests of the proposed external origin of the gas in early-type galaxies
will come from more detailed analysis of its spatial distribution
and kinematics. By studying the relation between gas and stellar
angular momentum in early-type galaxies, one can hope to gain further
understanding of how the gas was accreted.  Studies of this nature are planned
as part of the next generation of integral-field spectroscopy
studies of nearby early-type galaxies, e.g. the ATLAS3D survey
(http://www-astro.physics.ox.ac.uk/atlas3d/).

 \begin {figure*}
\begin {tabular}{cc}
\includegraphics[width=9cm]{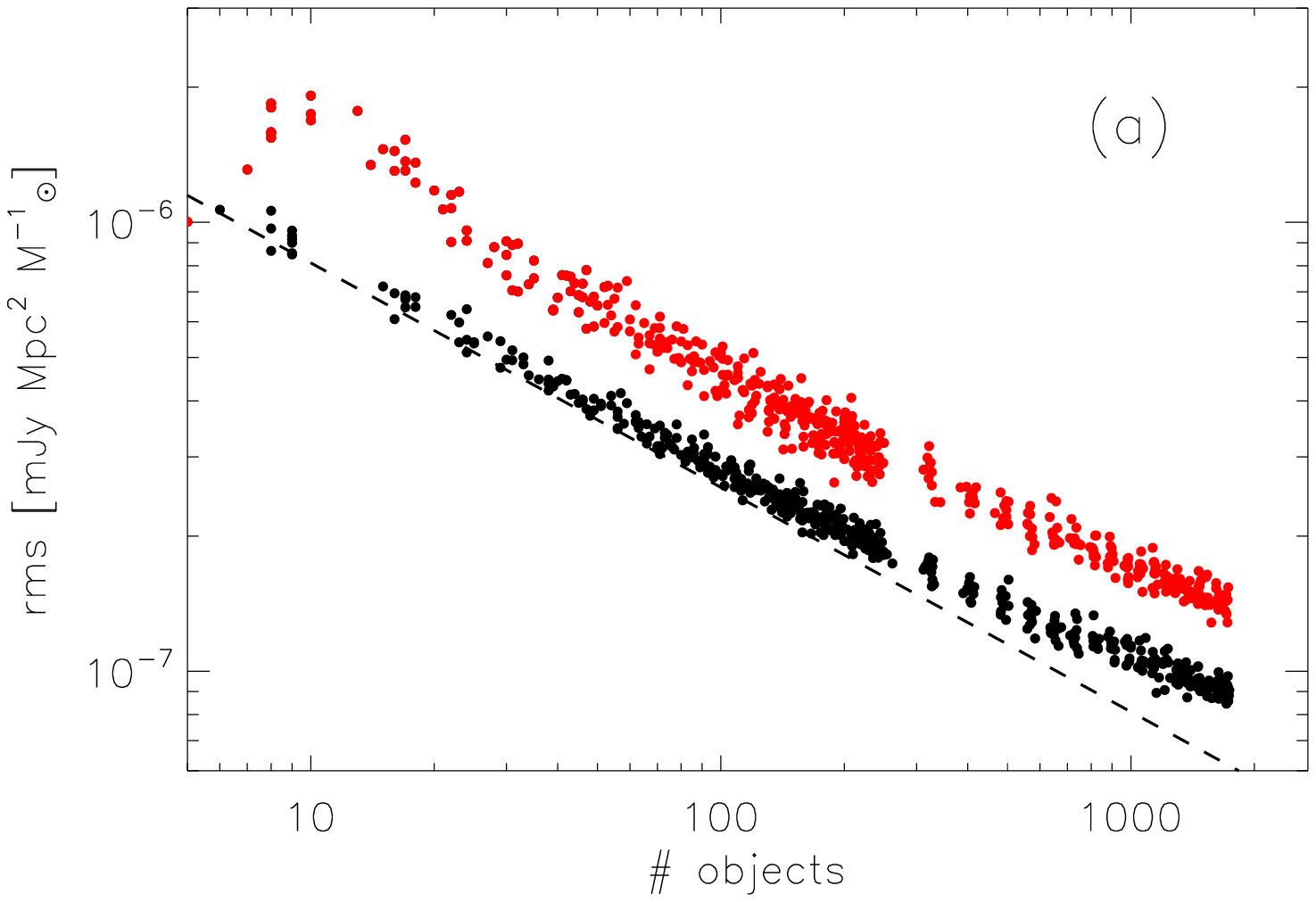}&
\includegraphics[width=8.cm]{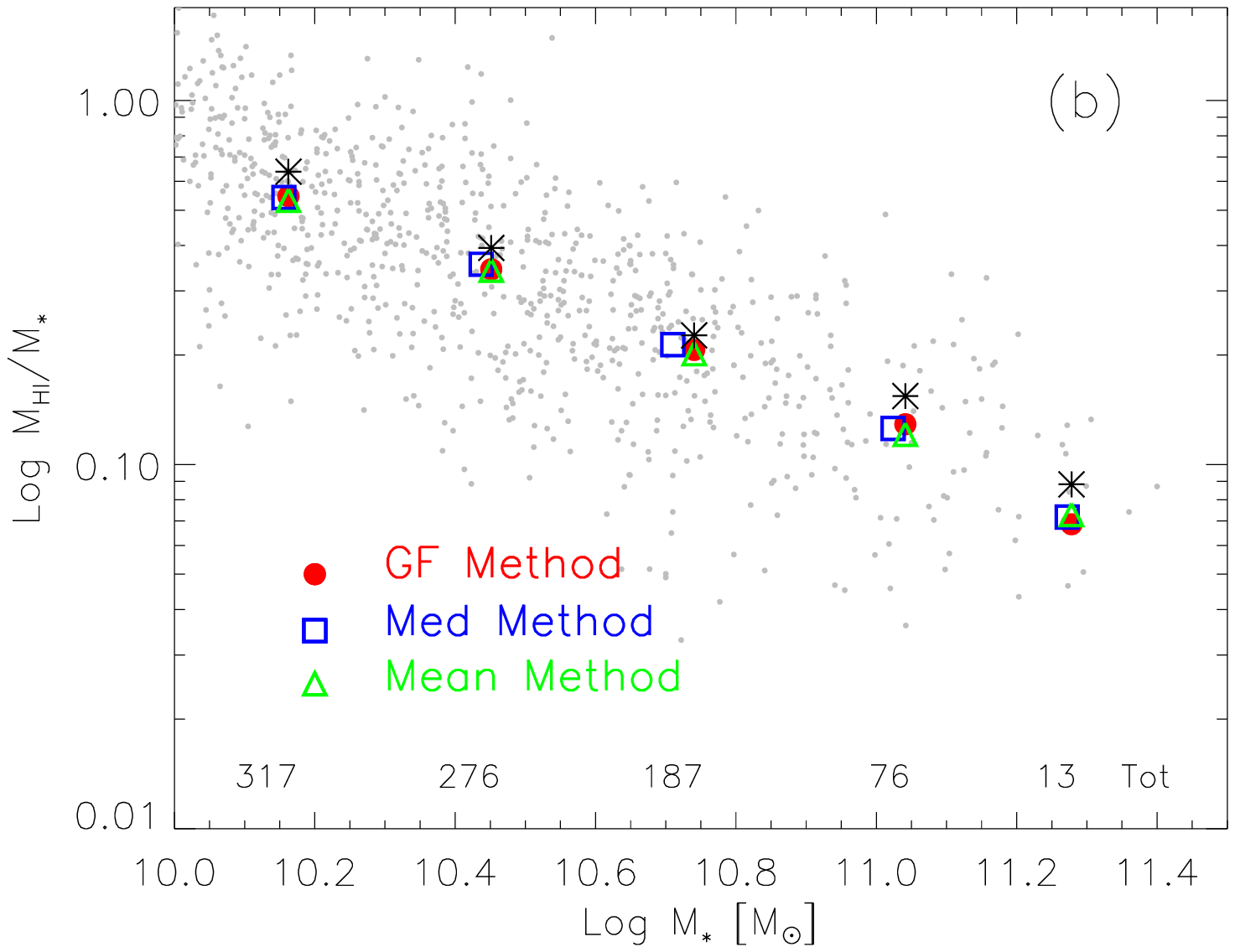}
\end {tabular}
\caption{(a) Dependence of the \textit{rms} of the stacked spectra on the 
number of objects stacked. The red dots are
obtained by stacking ``gas fractions''. The black ones by stacking
HI line fluxes. The dashed line is the expected 1/$\sqrt{N}$ dependence
(note that the black points have been multiplied by the average
squared distance and divided by the average {\Mst} in order to have the
same units as the red points [mJy$\,$Mpc$^2\,$M$^{-1}_{\odot}$]). 
(b) Comparison of average gas fractions obtained with the two different stacking
methods applied to \textit{sample A} galaxies with ALFALFA detections. 
Red circles are averages obtained using method 2, and green triangles/blue squares are from method (1), when mean
or median values of $z$ and $M_*$ are used in equation (A1),
respectively. The stars are the  means obtained from averaging the catalogued  
ALFALFA fluxes. Gray dots show the ALFALFA detections. The numbers of
objects co-added in each bin is reported below.}
\label{fig13}\end{figure*} 
\begin {figure*}
\begin {tabular}{cc}
\includegraphics[width=8.5cm]{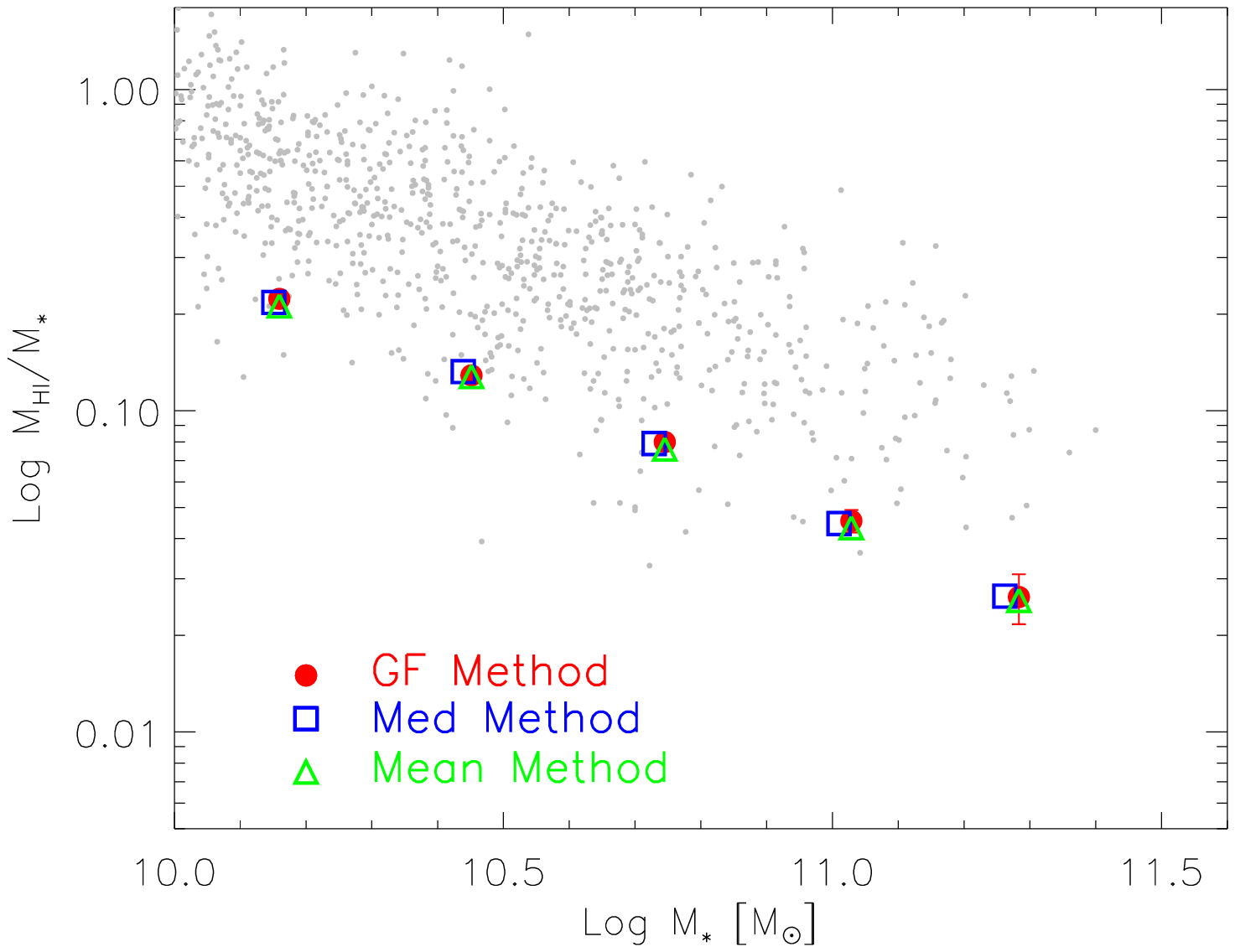} &
\includegraphics[width=8.5cm]{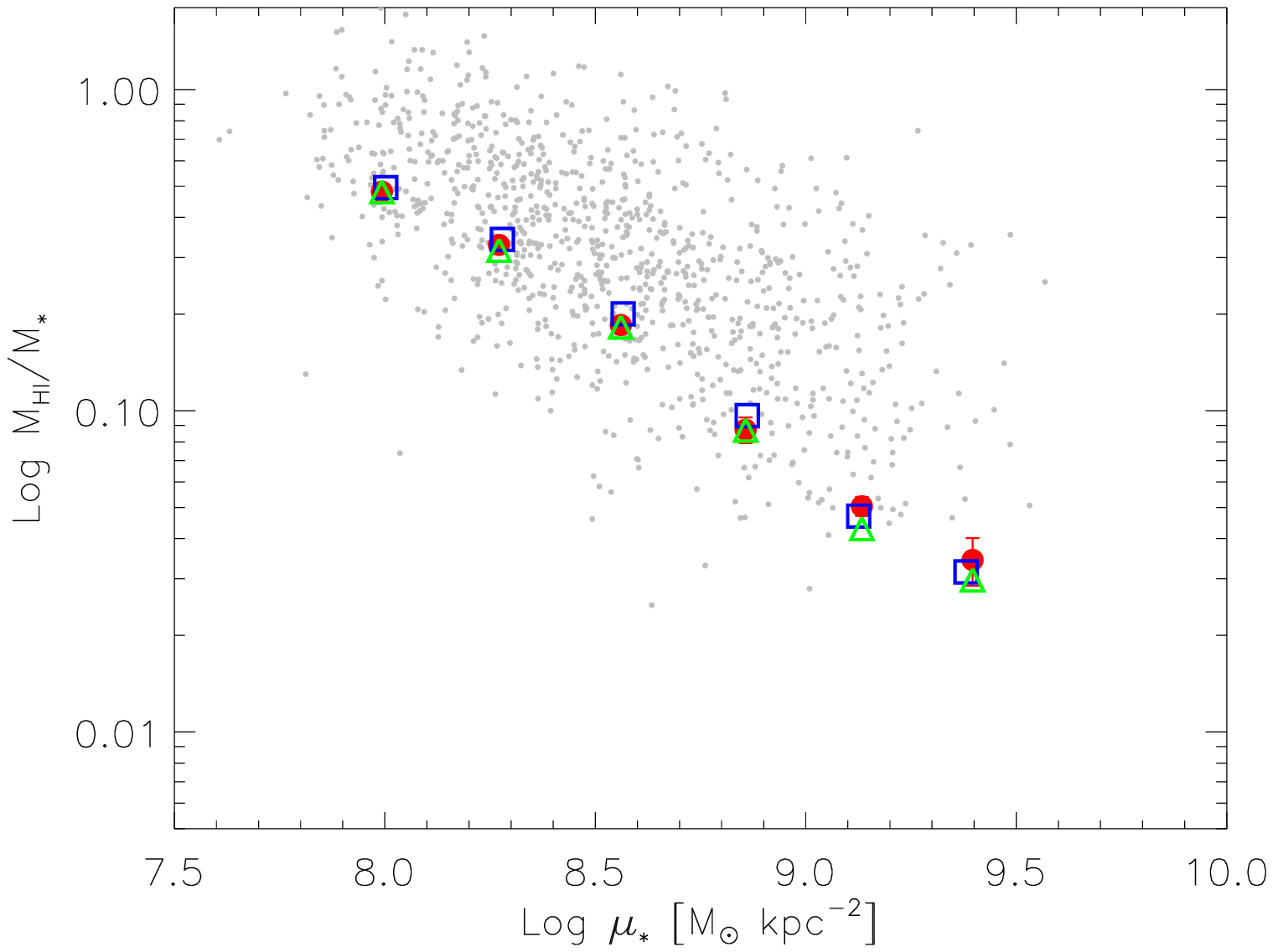}\\
\includegraphics[width=8.5cm]{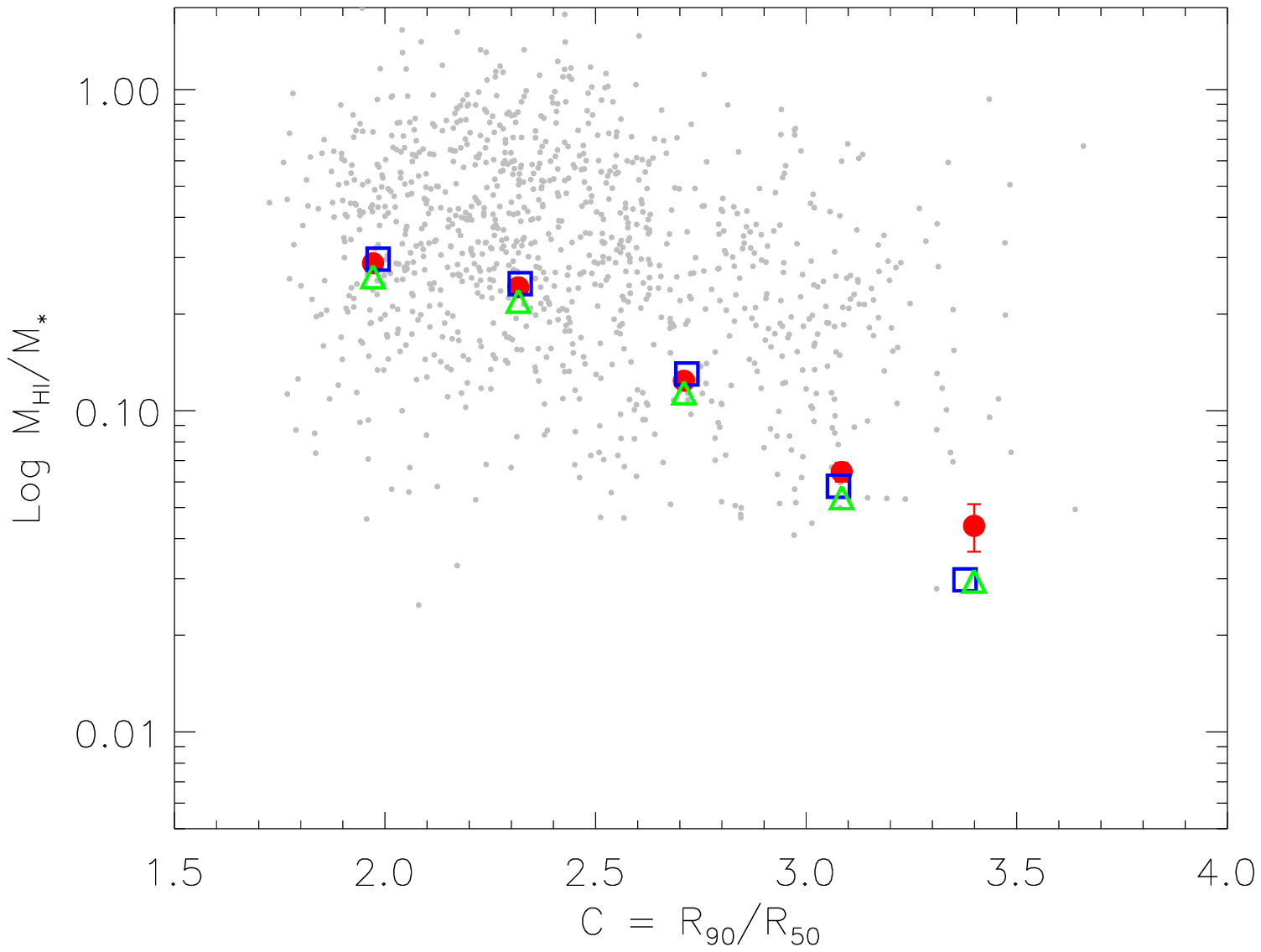} &
\includegraphics[width=8.5cm]{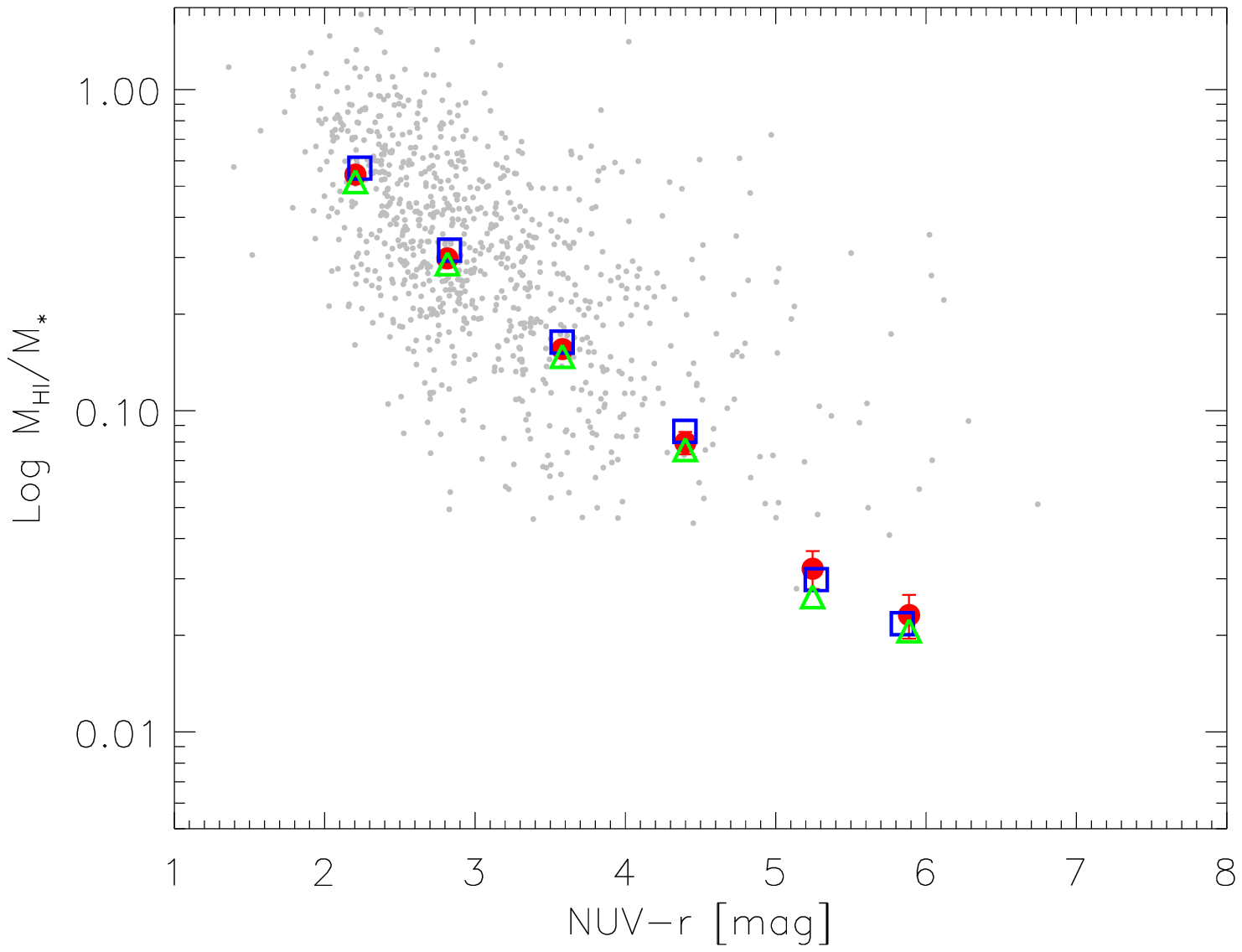}\\
\end {tabular}
\caption {Comparison of average gas fractions obtained with the two different stacking
methods, applied to all \textit{sample A} galaxies. 
Symbols and colours are the same as Figure \ref{fig13}.}
\label{fig14}
\end{figure*}

\appendix
\newcommand{\appsection}[1]{\let\oldthesection\thesection
  \section{#1}\let\thesection\oldthesection}

\appsection{Comparison of two different stacking procedures}

In this section we compare two different methods for deriving mean
{\hi} gas fractions from stacked spectra.\\

{\bf 1) Stacking {\hi} fluxes}. In the first method, we derive the average {\hi} gas fraction using the equation 
\begin{equation}\label{eq:gf_f}
\langle\frac{M_{HI}}{M_{*}}\rangle\,=\,\frac{2.356\times10^5}{1+\langle
  z\rangle}\frac{D^2_L(\langle z\rangle)}{\langle M_*\rangle}
\frac{\Sigma_{i=0}^N S_i\cdot w_i}{\Sigma_{i=0}^N w_i},
\end{equation}
where $\langle z \rangle$ is the mean redshift of the stacked galaxies and 
$\langle M_*\rangle $ is their mean stellar mass. The main limitation is that this method may produce biased
results if galaxies in the bin span a significant range in redshift or stellar mass.
By stacking  mainly
non-detected spectra, we do not know how much each galaxy  
contributes to the total signal. This approach will work best if the bins are small.
We note that our redshift range is compact enough so that: 
\begin{equation}D^2_L(\langle z\rangle) \simeq \langle D_L(z)\rangle^2 \simeq
\langle D^2_L(z).\rangle
\end{equation}
We choose to split each stellar mass  bin into three
sub-bins in redshift:
[0.025;0.033],[0.033;0.042],[0.417;0.050]. For each of these sub-bins, 
we  evaluate the mean stellar mass $\langle M_*\rangle$, and
split the sample into two further bins according to:
{\Mst}$\gtrless\langle M_*\rangle$. We stack the spectra in each of 
the 6 sub-bins, measure a flux and evaluate a gas fraction using both the average
and median values of $z$ and $M_*$. Then we evaluate the final gas fraction by averaging the six values obtained
(weighted by the number of objects co-added). \\

\par {\bf 2) Stacking ``gas fractions''}. This approach is 
described in section \ref{stacktool_b} and is the one that is adopted throughout the main body of this paper.\\

First, we checked that the $rms$ of the co-added
spectra decreases as 1/$\sqrt{N}$; when method (2)
is applied, i.e. by multiplying the flux by distance
and stellar mass, we are in fact rescaling  the
noise in the spectra in a sensible
way. Results are shown in 
Figure \ref{fig13} (panel a): red circles are
obtained by stacking ``gas fractions'', black circles are obtained by
stacking fluxes. The black points have been multiplied by the average
squared distance and divided by the average {\Mst} in order to have the
same units as the red points [mJy$\,$Mpc$^2\,$M$^{-1}_{\odot}$].
 The dashed line is the expected 1/$\sqrt{N}$ dependence,
which is still recovered  using method (2).

We then compared the results of stacking methods
(1) and (2). We divided all the galaxies in \textit{sample A} that
were detected by  ALFALFA  (detection codes 1+2)  into five
stellar mass bins. We compare results from the different stacking
methods with the results obtained by averaging together the individual 
catalogued measurements. The results are
shown in Figure \ref{fig13} (panel b), where
red circles show results obtained using method
(2), and green triangles/blue squares are from method (1), when mean
or median values of $z$ and $M_*$ are used in equation (A1),
respectively. The stars show the results from averaging together the
individual detections. Gray dots show the ALFALFA detections.
The signal recovered from the stacking is consistent with the mean
value of the individual detections for each bin.  The two different stacking methods yield results that are
also consistent with each other. Small differences of around $\sim 10$ \% in
{\Mhi}/$M_*$  do occur in the two largest mass bins that
contain the fewest objects. \\
We  also compared results for the  different methods including the non-detections.
In Figure \ref{fig14} we show the same correlations studied in section
\ref{stackGass} for \textit{sample A}, computed using different  methods. Symbols
are the same as described above. Once again, we obtain good consistency.                   

\section*{Acknowledgments}
SF wishes to thank Genevieve J. Graves and Greg Hallenbeck for useful discussions.
We thank the many members of the ALFALFA team who have contributed to the acquisition
and processing of the ALFALFA dataset over the last six years.

RG and MPH are supported by NSF grant AST-0607007 and by a grant from the Brinson
Foundation.

The Arecibo Observatory is part of the National Astronomy and
Ionosphere Center, which is operated by Cornell University under a
cooperative agreement with the National Science Foundation.

GALEX is a NASA Small Explorer, launched in 2003 April. We gratefully
acknowledge NASA's support for construction, operation and science
analysis for the GALEX mission, developed in cooperation with the
Centre National d'Etudes Spatiales (CNES) of France and the Korean
Ministry of Science and Technology.

Funding for the SDSS and SDSS-II has been provided by the Alfred
P. Sloan Foundation, the Participating Institutions, the National
Science Foundation, the U.S. Department of Energy, the National
Aeronautics and Space Administration, the Japanese Monbukagakusho, the
Max Planck Society and the Higher Education Funding Council for
England. The SDSS web site is http://www.sdss.org/. 

The SDSS is managed by the Astrophysical Research Consortium for the
Participating Institutions. The Participating Institutions are the
American Museum of Natural History, Astrophysical Institute Potsdam,
University of Basel, University of Cambridge, Case Western Reserve
University, University of Chicago, Drexel University, Fermilab, the
Institute for Advanced Study, the Japan Participation Group, Johns
Hopkins University, the Joint Institute for Nuclear Astrophysics, the
Kavli Institute for Particle Astrophysics and Cosmology, the Korean
Scientist Group, the Chinese Academy of Sciences (LAMOST), Los Alamos
National Laboratory, the Max-Planck-Institute for Astronomy (MPIA),
the Max-Planck-Institute for Astrophysics (MPA), New Mexico State
University, Ohio State University, University of Pittsburgh,
University of Portsmouth, Princeton University, the United States
Naval Observatory and the University of Washington.

\begin {table*}
\begin{tabular}{|cc|c|cc|cc|cc|cc|}
\hline
 & & & $\langle\,${\Mhi}/{\Mst}$\rangle$ & &
$\langle\,${\Mhi}/{\Mst}$\rangle$ & &$\langle\,${\Mhi}/{\Mst}$\rangle$&
& $\langle\,${\Mhi}/{\Mst}$\rangle$
&  \\  
$x$& $\langle\,x\rangle$ & $a$ &\emph{Sample A} & N & \emph{ETG sample} &  N
& \emph{ETG sample 1} & N & \emph{ETG sample 2} & N     \\
 \hline
Log {\Mst} & 10.16 &$-0.78\,\pm
\,0.03$ &0.222$\pm$0.008 & 1734 & 0.056$\pm$0.009 & 493 &
0.091$\pm$0.011 & 311 & 0.118$\pm$0.019 & 139\\
    & 10.45 & &0.128$\pm$0.005 & 1538 & 0.036$\pm$0.005 & 566 &
0.056$\pm$0.006 & 389 & 0.049$\pm$0.006 & 294\\
    & 10.74 & &0.079$\pm$0.004 & 1025 & 0.033$\pm$0.004 & 516 &
0.043$\pm$0.005 & 351 & 0.041$\pm$0.005 & 328\\
    & 11.03 & &0.044$\pm$0.004 & 430  & 0.015$\pm$0.004 & 246 &
0.018$\pm$0.003 & 183 & 0.014$\pm$0.004 & 178\\
    & 11.28 & &0.026$\pm$0.005 & 63   & 0.016$\pm$0.004 & 43  &
0.007$^*$ & 32  & 0.009$^*$ & 25 \\
\hline
 $C$ & 1.97 &$-0.59\,\pm\,0.03$ &0.288$\pm$0.013 & 420 &-&-&-&-&-&-\\
 $(R_{90}/R_{50})$ & 2.32 && 0.242$\pm$0.008 & 1149 &-&-&-&-&-&-\\
    & 2.71 && 0.124$\pm$0.006 & 1464 &-&-&-&-&-&-\\
    & 3.08 && 0.064$\pm$0.004 & 1409 &-&-&-&-&-&-\\
    & 3.40 && 0.044$\pm$0.007 & 287 &-&-&-&-&-&-\\
\hline
Log {\must} & 7.99 &$-0.87\,\pm
\,0.03$ &0.481$\pm$0.025 & 193 & - & - & - & - & - & - \\
   & 8.27 & &0.328$\pm$0.011 & 625  & 0.293$\pm$0.082 & 8   &
0.336$\pm$0.109 & 5   & 0.435$\pm$0.154 & 3\\
   & 8.56 & &0.185$\pm$0.008 & 1003 & 0.112$\pm$0.018 & 148 &
0.107$\pm$0.019 & 120 & 0.139$\pm$0.047 & 41\\
   & 8.86 & &0.087$\pm$0.008 & 1371 & 0.077$\pm$0.006 & 661 &
0.071$\pm$0.007 & 461 & 0.073$\pm$0.010 & 258\\
   & 9.13 & &0.050$\pm$0.003 & 1237 & 0.043$\pm$0.004 & 857 &
0.037$\pm$0.004 & 591 & 0.036$\pm$0.004 & 535\\
   & 9.40 & &0.034$\pm$0.006 & 287  & 0.035$\pm$0.008 & 183 &
0.029$\pm$0.011 & 89 & 0.020$\pm$0.005 & 124\\
\hline
 {\col} & 2.20 &$-0.38\,\pm \,0.01$ &0.543$\pm$0.024 & 209 & 0.351$\pm$0.091 & 14
 & 0.301$\pm$0.121 & 10  & 0.569$\pm$0.293 & 3\\
   & 2.82 & &0.298$\pm$0.009 & 855 & 0.253$\pm$0.019 & 123
 & 0.220$\pm$0.020 & 77  & 0.247$\pm$0.038 & 29\\
   & 3.58 & &0.156$\pm$0.007 & 760 & 0.127$\pm$0.011 & 211
 & 0.125$\pm$0.014 & 127 & 0.152$\pm$0.020 & 74\\
   & 4.40 & &0.079$\pm$0.006 & 609 & 0.073$\pm$0.009 & 278
 & 0.072$\pm$0.013 & 180 & 0.084$\pm$0.011 & 118\\
   & 5.24 & &0.032$\pm$0.004 & 909 & 0.032$\pm$0.005 & 669
 & 0.027$\pm$0.005 & 468 & 0.022$\pm$0.004 & 387\\
   & 5.89 & &0.023$\pm$0.004 & 621 & 0.024$\pm$0.004 & 490
 & 0.014$\pm$0.004 & 349 & 0.021$\pm$0.004 & 315\\
\hline
Log $\sigma$ &1.90& -& 0.149$\pm$0.014 & 287 & 0.077$\pm$0.020 & 76  &
0.085$\pm$0.026 & 50 & 0.058$\pm$0.016 & 99\\
  & 2.01 & &0.135$\pm$0.009 & 483 & 0.089$\pm$0.014 & 169 &
0.074$\pm$0.016 & 114 & 0.059$\pm$0.006 & 302\\
  & 2.10 & &0.077$\pm$0.005 & 738 & 0.054$\pm$0.007 & 353 &
0.047$\pm$0.008 & 234 & 0.059$\pm$0.010 & 136\\
  & 2.20 & &0.053$\pm$0.005 & 711 & 0.041$\pm$0.004 & 439 &
0.042$\pm$0.006 & 287 & 0.048$\pm$0.006 & 252\\
  &2.29 & &0.023$\pm$0.003 & 436 & 0.022$\pm$0.003 & 322 &
0.022$\pm$0.004 & 222 & 0.025$\pm$0.004 & 246\\
  & 2.40 && 0.010$\pm$0.002 & 167 & 0.012$\pm$0.002 & 141 &
0.008$^*$ & 112 &0.009$^*$ & 123\\
\hline
\end{tabular}\caption{Average gas fractions for the samples shown in
  Figures \ref{fig07}-\ref{fig10}: \textit{sample A} (Fig. \ref{fig07} and
  \ref{fig08}, red circles), the main \textit{ETG
    sample} (Fig. \ref{fig09} and
  \ref{fig10}, blue circles), and the two \emph{ETG} sub-samples defined with
  more stringent cuts on {\cix} and inclination (Fig. \ref{fig09} and
  \ref{fig10}, black and cyan circles). For \textit{sample
    A} we also report the slopes $a$ (third column) of the relations
  $\log\,\langle\,${\Mhi}/{\Mst}$\rangle\,\,=a\langle\,x\rangle+b$,
  where $x$ is the quantity listed in the 1st column.}\label{tab01}
\end{table*}

\end{document}